\documentclass[aps,
pra,
twocolumn,
reprint,
superscriptaddress,
nofootinbib]{revtex4-2}

\usepackage[svgnames,table,xcdraw]{xcolor}
\usepackage{amsmath}
\usepackage{microtype}
\usepackage{amssymb}
\usepackage{amsmath}
\usepackage{braket}
\usepackage{graphicx}
\usepackage{booktabs}
\usepackage{tabu}
\usepackage{bm}
\usepackage{dsfont}
\usepackage{hyperref}
\usepackage[caption=false]{subfig}
\usepackage{amsthm}
\newtheorem*{claim}{Claim}
\usepackage{xfrac}
\hypersetup{colorlinks,linkcolor={DarkBlue},citecolor={DarkBlue},urlcolor={DarkBlue}}

\let\originalleft\left
\let\originalright\right
\renewcommand{\left}{\mathopen{}\mathclose\bgroup\originalleft}
\renewcommand{\right}{\aftergroup\egroup\originalright}

\usepackage{nag}
\usepackage{graphicx}
\usepackage{amsthm}
\usepackage{tabulary}
\usepackage{qcircuit}
\usepackage{mathtools}
\usepackage{bm}
\usepackage{braket}
\usepackage{datetime}
\usepackage{stackrel}
\usepackage{dsfont}
\usepackage{qcircuit}
\usepackage[english]{babel}
\usepackage{units}

\usepackage{tikz}
\usetikzlibrary{backgrounds,decorations.pathreplacing}

\usepackage{chngcntr}
\counterwithout{equation}{section}

\usepackage[normalem]{ulem}
\usepackage{slashed}
\usepackage{soul}

\usepackage{xcolor}
\usepackage{amssymb}
\usepackage{amsmath}
\usepackage{amsthm}
\usepackage{braket}
\usepackage{mathdots}

\usepackage{makeidx}
\makeindex

\usepackage{soul}
\usepackage{xargs}
\newcommandx{\cmnote}[2][1=]{\linespread{1.0}\todo[linecolor=red,backgroundcolor=red!25,bordercolor=red,#1]{#2}}

\let\underline\ul

\allowdisplaybreaks[4]

 \index{} \index{} \index{} \index{} \index{} \index{} \index{} \index{}%%
\makeatletter

\newcommand{\ringplus}{\mathbin{\text{\@ringplus}}}

\newcommand{\@ringplus}{%
  \ooalign{\hidewidth\raise1.3ex\hbox{\tiny$\circ$}\hidewidth\cr$\m@th+$\cr}%
}

\newcommand{\ringminus}{\mathbin{\text{\@ringminus}}}

\newcommand{\@ringminus}{%
  \ooalign{\hidewidth\raise0.9ex\hbox{\tiny$\circ$}\hidewidth\cr$\m@th-$\cr}%
}
\makeatother
 \index{} \index{} \index{} \index{} \index{} \index{} \index{} \index{}%%

\newcommand{\tp}[0]{T}

\newcommand{\qunaught}{\varnothing}

\newcommand{\latG}{V}
\DeclareFontFamily{U}{wncy}{}
\DeclareFontShape{U}{wncy}{m}{n}{<->wncyr10}{}
\DeclareSymbolFont{mcy}{U}{wncy}{m}{n}
\DeclareMathSymbol{\Sh}{\mathord}{mcy}{"58}

\newcommand{\negspace}{\!}
\newcommand{\lsub}[2]{{\protect\vphantom{#1}}_{#2} \negspace {#1}}
\newcommand{\rsub}[2]{{#1} \negspace {\protect\vphantom{#1}}_{#2}}

\newcommand{\ketsub}[2]{\rsub {\ket{#1}} {#2}}
\newcommand{\brasub}[2]{\lsub {\bra{#1}} {#2}}

\newcommand{\pbra}[1]{\brasub{#1} p}
\newcommand{\qbra}[1]{\brasub{#1} q}
\newcommand{\pket}[1]{\ketsub{#1} p}
\newcommand{\qket}[1]{\ketsub{#1} q}

\newcommand{\cent}{C}
\newcommand{\sat}{S}

\newcommand{\op}[1]{\hat{#1}}

\newcommand{\opvec}[1]{\op{\vec{#1}}}

\newcommand{\mat}[1]{\bm{\mathrm{#1}}}

\renewcommand{\vec}[1]{\bm{#1}}

\newcommand{\controlled}[1]{\op{\mathrm{C}}_{#1}}
\newcommand{\CZ}[0]{\controlled Z}
\newcommand{\CX}[0]{\controlled X}

 \index{} \index{} \index{} \index{} \index{} \index{} \index{} \index{}%%
 \index{} \index{} \index{} \index{} \index{} \index{} \index{} \index{}%%

 \index{} \index{} \index{} \index{} \index{} \index{} \index{} \index{}%%
 \index{} \index{} \index{} \index{} \index{} \index{} \index{} \index{}%%
\usepackage{graphicx}
\usepackage{graphicx}
\usepackage{multirow}
\usepackage{hhline}
\xyoption{color}
\xyoption{line}

\newcommandx*\bsbal[3][1=black, 3=->]{\ar @[#1]@{#3} [#2,0] \qw}

\newcommandx*\varbs[5][1=black, 3=\theta,4=0.5,5=->]{\ar @[#1]@{#5}^(#4){#3} [#2,0] \qw}

\newcommand{\ctrlnowire}[1]{\control \qwx[#1]}
\newcommandx*\lblline[3][3=0.5]{\ar @{-}^(#3){#1} [#2,0]}
\newcommandx*\ctrlg[3][3=0.5]{ \raisebox{-3pt}{$\bullet$}  \ar @{-}^(#3){#1} [#2,0] \qw }
\newcommandx*\ctrlog[2]{\controlo \ar @{-}^{#1} [#2,0] \qw}
\newcommandx*\ctrlodash[1]{\controlo \ar @{-} [#1,0] \ar @[black]@{.} [0,-1]}
\newcommand{\qwdot}{\ar @[black]@{.} [0,-1]}
\newcommand{\nogate}[1]{*+<.6em>{#1} \POS ="i","i"+UR;"i"+UL **[white]\dir{-};"i"+DL **[white]\dir{-};"i"+DR **[white]\dir{-};"i"+UR **[white]\dir{-},"i" }

\newcommand{\nomultigate}[2]{*+<1em,.9em>{\hphantom{#2}} \POS [0,0]="i",[0,0].[#1,0]="e",!C *{#2},"e"+UR;"e"+UL **[white]\dir{-};"e"+DL **[white]\dir{-};"e"+DR **[white]\dir{-};"e"+UR **[white]\dir{-},"i"}

\begin{document}

\title{Linear-optical quantum computation with arbitrary error-correcting codes}

\def \affXanadu {Xanadu Quantum Technologies Inc., Toronto ON, Canada}
\def \affRMIT {Centre for Quantum Computation and Communication Technology, School of Science, RMIT University, Melbourne, VIC 3000, Australia}
\def \affMacquarie {Center for Engineered Quantum Systems, School of Mathematical
and Physical Sciences, Macquarie University, Sydney, NSW 2109, Australia}
\def \affPI {Perimeter Institute for Theoretical Physics, Waterloo, ON N2L 2Y5, Canada}
\def \affWaterloo {Institute for Quantum Computing, University of Waterloo, Waterloo, ON N2L 3G1, Canada}

\author{Blayney W. Walshe}
\affiliation{\affXanadu}
\author{Ben Q. Baragiola} 
\affiliation{\affXanadu}
\affiliation{\affRMIT}
\author{Hugo Ferretti}
\affiliation{\affXanadu}
\author{Jos\'{e} Gefaell}
\affiliation{\affXanadu}
\author{Michael Vasmer}
\affiliation{\affXanadu}
\affiliation{\affPI}
\affiliation{\affWaterloo}
\author{Ryohei Weil}
\affiliation{\affXanadu}
\author{Takaya Matsuura}
\affiliation{\affXanadu}
\affiliation{\affRMIT}
\author{Thomas Jaeken}
\affiliation{\affXanadu}
\author{Giacomo Pantaleoni}
\affiliation{\affXanadu}
\affiliation{\affMacquarie}
\author{Zhihua Han}
\affiliation{\affXanadu}
\author{Timo Hillmann}
\affiliation{\affXanadu}
\author{Nicolas C. Menicucci}
\affiliation{\affXanadu}
\affiliation{\affRMIT}
\author{Ilan Tzitrin}
\email{ilan@xanadu.ai}
\affiliation{\affXanadu}
\author{Rafael N. Alexander}
\affiliation{\affXanadu}

\date{\today}

\begin{abstract}
\end{abstract}

\begin{abstract}
High-rate quantum error correcting codes mitigate the imposing scale of fault-tolerant quantum computers but require efficient generation of non-local, many-body entanglement. We provide a linear-optical architecture with these properties, compatible with arbitrary codes and Gottesman-Kitaev-Preskill qubits on generic lattices, and featuring a natural way to leverage physical noise bias. Simulations of hyperbolic surface codes and bivariate bicycle codes, promising families of quantum low-density parity-check codes, reveal a threshold comparable to the 2D surface code with substantially better encoding rates.
\end{abstract}

\maketitle

\section{Introduction}

Promising quantum algorithms often require millions of physical qubits due to the overheads for performing quantum error correction (QEC), commensurate with projected levels of noise across physical systems~\cite{Gidney2021}.
A significant reduction in the resource costs for fault-tolerant quantum computation is possible with the use of quantum low-density parity check (qLDPC) codes with high encoding rates~\cite{breuckmann2021lqdpc,panteleev2022, leverrier2022, dinur2022, Bravyi2024}. However, practical implementations of these codes are hindered by the requirement of non-geometrically-local entanglement, often across many qubits due to the presence of high-weight stabilizers. 

Bolstered by optical fibres, photonic platforms natively support arbitrary connectivity between qubits, making them ideally suited for implementing qLDPC codes. However, optical architectures can vary significantly in their properties; for example, some cannot generate entanglement deterministically, impacting their scalability~\cite{Kok2007,slussarenko2019photonic}. Approaches based on optical Gottesman-Kitaev-Preskill (GKP) qubits~\cite{GKP}, in which Clifford gates map to Gaussian operations, face no such restriction, benefiting from deterministic entanglement generation as well as fast gates and measurements at room temperature and pressure. Still, existing GKP-based architectures have either demanded noisy hardware components---inline squeezers and fast switches~\cite{bourassa2021blueprint}---or been restricted to the 2D surface code, which has a vanishing encoding rate~\cite{tzitrin2021fault}.

In this work, we resolve the limitations of the above approaches through a ``tale of two lattices'':
a fault-tolerant quantum computing architecture capable of generating non-local many-body entanglement deterministically with static linear optics for any choice of quantum error-correcting code. We show furthermore that the architecture is compatible with GKP states defined on any lattice and provide a way to leverage the resulting anisotropic noise. This accommodates state preparation approaches that are naturally constrained to generating non-square GKP states or those that can be tuned to produce these states for noise engineering. This treatment allows us to decouple the primal and dual decoding problems, which we use in simulations involving two promising families of qLDPC codes: hyperbolic surface codes~\cite{freedman2002z2,Breuckmann_2016} and bivariate bicycle codes~\cite{Bravyi2024}. For both, we observe comparable thresholds to the 2D surface code with substantial improvements to the encoding rate.

\section{Arbitrary graph states from passive transformations on GKP states}

Photonic implementations of quantum computers require low-depth optical circuits. They operate best in the paradigm of measurement-based quantum computing (MBQC), where a few layers of a multi-partite entangled state called a cluster or graph state~\cite{hein2006} are prepared first, and then the gates are performed by way of adapative local measurements on this state~\cite{Raussendorf2001, Raussendorf2003}. Quantum error correction circuits readily translate to the measurement-based model, where parity checks become combinations of graph-state stabilizer measurements~\cite{Bolt2016, hillmann2024}.

\begin{figure}
\subfloat[
]{\label{subfig:qubit_graph}\includegraphics[width=.4\columnwidth]{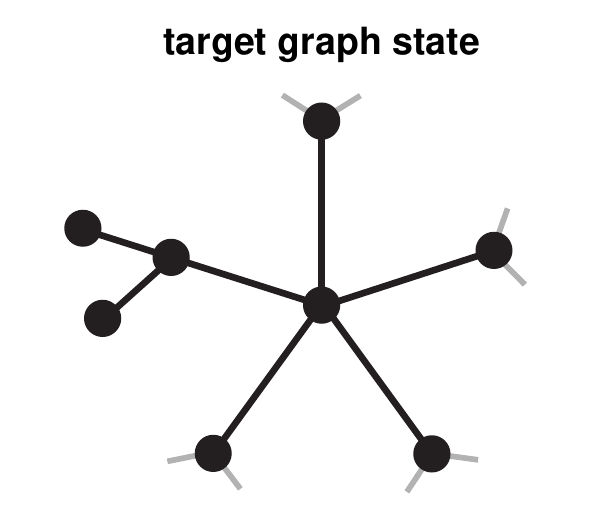}}
\subfloat[
]{\label{subfig:macro_graph}\includegraphics[width=.4\columnwidth]{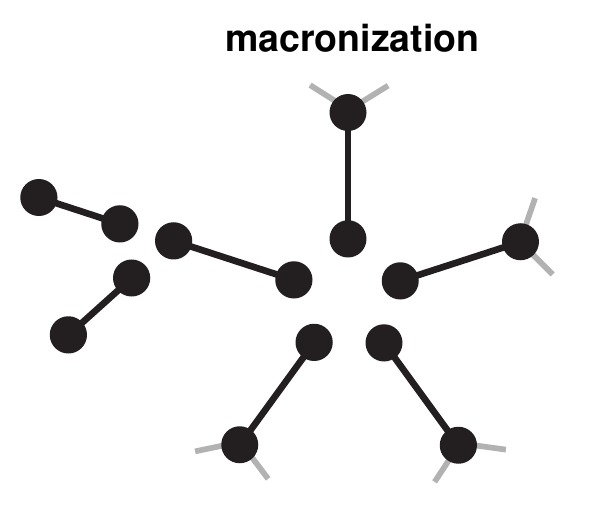}}

\subfloat[
]{\label{subfig:stitching}\includegraphics[width=.4\columnwidth]{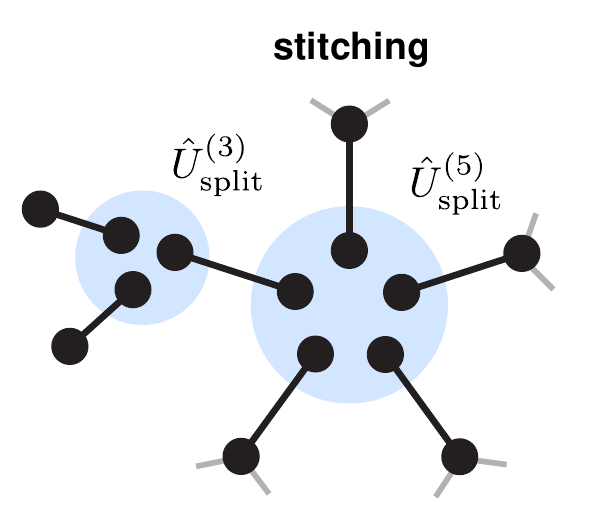}} 
\subfloat[
]{\label{subfig:reduction}\includegraphics[width=.4\columnwidth]{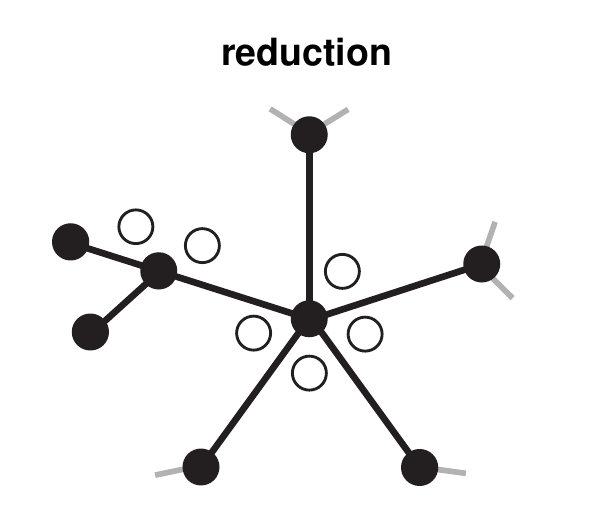}}
    \caption{General procedure for constructing a qubit graph state using GKP dumbbell states and passive linear optics. 
    (a) Example target qubit graph state.  
    (b) Replace each edge in the graph with a GKP dumbbell. 
    (c) Apply a splitter (blue circles) to each macronode, \emph{i.e.} the modes meeting at each node of the qubit graph.
    (d) Measure $n-1$ satellite modes (hollow circles) at each node to distentangle them, producing the desired GKP graph state up to local Gaussian unitaries.}\label{fig:dumbbell_splitter_method}
\end{figure}

A measurement-based quantum information task, such as a logical measurement on an encoded qubit or the transmission of quantum information to distant parties, amounts to a specification of a target qubit graph state $\Ket{G}$ (such as the one in Fig.~\ref{subfig:qubit_graph}) with underlying graph $G$, along with a sequence of local measurements on $\Ket{G}$ and classical feedforward. We call $G$ the \emph{canonical} (reduced) graph, the target of our study. The nodes and edges of $G$ are associated, respectively, with qubit $\Ket{+}$ states and controlled-Z (CZ) gates. The \emph{valence} of a node refers to the number of its neighbors or incident edges; for a QEC code, this is related to the number of qubits in a parity check or the number of checks a qubit participates in. 

Here we give the necessary ingredients---summarized in Fig.~\ref{fig:dumbbell_splitter_method}---for creating and measuring any target graph state $\Ket{G}$ given three types of resource: GKP states, passive Gaussian transformations (phase shifters and beam splitters), and homodyne measurements.

\emph{Macronization}---%
In the first step of our protocol, Fig.~\ref{subfig:macro_graph}, we associate each edge of $G$ with a \emph{GKP dumbbell state} constructed unitarily by entangling two qunaught states $\ket{\varnothing} \coloneqq \sum_{k\in \mathbb{Z}} \qket{\sqrt{2\pi}k}$~\cite{Walshe2020}, 
\begin{align}\label{eq:dumbbell}
    \begin{split}
    \Qcircuit @C=0.3cm @R=0.3cm
    {
    & \nogate{\ket{\qunaught}} &\bsbal{1} &\qw      &\qw \\
    & \nogate{\ket{\qunaught}} &\qw       &\gate{F} 
    &\qw
    }
    \quad \raisebox{-0.45cm}{=} \,
    \Qcircuit @C=0.5cm @R=0.3cm
    {
    & \nogate{\Ket{+}} & \ctrl{1} & \qw \\
    & \nogate{\Ket{+}} & \ctrl{0} & \qw
    }
    \quad \raisebox{-0.45cm}{=} \,
    \Qcircuit @C=0.5cm @R=0.78cm
    {
     & \ctrlnowire{1} \\
     & \ctrlnowire{0}
    }
    \end{split}
\end{align}
where the arrow denotes a real beam splitter, $\op B_{jk}(\theta)\coloneqq e^{-i\theta(\op q_j \op p_k - \op p_j \op q_k)}$. In Eq.~\eqref{eq:dumbbell}, the beam splitter is $\emph{balanced}$, meaning its transmission angle is $\theta = \frac{\pi}{4}$ (in this special case we do not specify the angle on the diagram). The remaining gates in the circuit act logically in the square-lattice GKP qubit space: $\op{F} \coloneqq e^{i\frac{\pi}{2}\op{n}}$ is a Fourier transform that behaves as a qubit Hadamard gate, and the two-mode gate $\op{C}_Z \coloneqq e^{i \op{q} \otimes \op{q}}$ is a weight-1 CV CZ gate that performs a qubit CZ on the GKP $X$ eigenstate $\ket{+} \coloneqq \sum_{k \in \mathbb{Z}} \Ket{\sqrt{\pi} k}_q$.\footnote{The input states can also be replaced with some combination of position and momentum squeezed states (in case of probabilistic sources) and GKP magic states (for non-Clifford operations)~\cite{tzitrin2021fault}.}

Through this mapping, each valence-$n$ node of $G$ is associated with $n$ dumbbell halves forming a multimode site called a \emph{macronode}~\cite{alexander2016flexible, Alexander2014,tzitrin2021fault}, indicated by blue circles in Fig.~\ref{subfig:stitching}. Physically, dumbbells can be created by sending two qunaught states generated on a photonic chip~\cite{Tzitrin2020, bourassa2021blueprint, takase2022gaussian, madsen2022quantum, Takase2024, Konno2024} through a directional coupler. Then, the dumbbell halves can be routed via optical fibres to spatially separate locations corresponding to each macronode. Additional time delays can be introduced for a subset of the modes to extend the computation in time~\cite{tzitrin2021fault}.

\emph{Stitching and reduction}---%
The modes at each macronode must be \emph{stitched} (entangled) together.\footnote{In Ref.~\cite{tzitrin2021fault}, where the target graph state is the RHG lattice, the macronodes---containing four modes in the bulk of the lattice---are stitched with a balanced four-splitter~\cite{Walshe2023equivalent}. For valences smaller than four, such as on the boundaries, the authors of~\cite{tzitrin2021fault} suggest introducing and then decoupling ancillary modes.} A general quantum information task, such as a fault-tolerant logical operation, may require a graph state with any valence, even in light of recent advances in stabilizer weight reduction~\cite{sabo2024}. To accommodate graph states with arbitrary connectivity, we introduce a class of linear-optical unitaries---decomposable into real beam splitters---that we refer to generically as \emph{splitters}. A splitter $\op{U}^{(n)}_\text{split}$ serves to entangle the $n$ modes at a single macronode (Fig.~\ref{subfig:stitching}), which are partitioned into a \textit{central} mode and $n-1$ \textit{satellite} modes. Measuring the satellite modes in $\op{q}$ disentangles them and leaves behind the central mode to serve as a node in the desired graph state (Fig.~\ref{subfig:reduction}).
By performing this process at each macronode, one obtains a graph between the (potentially) unmeasured central modes, connected to each other by weight-1 CZ gates, up to byproduct Gaussian operations---local squeezing and displacements---that can be accounted for in post-processing. We call this the \emph{reduction} to the canonical graph state $\ket{G}$.

The working principle behind stitching and reduction is the equivalence, under measurement of the satellite modes, between a splitter and a star-type network of weight $-1$ CV controlled-X (CX) gates, each of the form $\op{C}_X \coloneqq e^{i \op{q} \otimes \op{p}}$. The CZ gates from all dumbbells meeting at a macronode are copied by this network onto the central mode, producing the desired connectivity there. Previously, this relationship was known only for a four-mode beam splitter network called a balanced four-splitter~\cite{Walshe2023equivalent}, limiting the construction to valence-4 graph states, such as the RHG lattice~\cite{tzitrin2021fault}. In the Supplemental Material, we detail several families of splitters that work over any number of modes, enabling the construction of all measurement-based QEC codes or any other graph states. Different families feature different beam-splitter arrangements and transmissivities, providing flexibility for experimentalists to choose the best splitter given laboratory constraints.

For an explicit example of a family of splitters and their role in stitching and reduction, consider the  $\emph{cascade splitters}$. In this family, splitters couple adjacent modes according to $\hat{U}^{(n)}_\text{cascade} = \prod_{k=1}^{n-1} \hat{B}_{n-k+1,n-k}(\theta_{k})$,
with beam splitters of lower $k$ acting first, and the transmission angles are tuned to $\theta_{k} = -\tan^{-1} \sqrt{k}$. This gives a depth $n-1$ beam-splitter network, where each mode interacts with at most two beam splitters.
Using a cascade splitter in the five-mode macronode in Fig.~\ref{fig:dumbbell_splitter_method} and measuring the satellite modes (with 0-measurement outcomes for simplicity) gives
\begin{equation}\label{bsNetworkcascade}
\begin{split}
\resizebox{0.91\columnwidth}{!}{
       \!\!\!\!\!\Qcircuit @C=0.4cm @R=0.13cm
        {
          &\qw  &\qw    &\qw    &\varbs{1}[\theta_{4}] & \qw & \nomultigate{4}{= \quad} &\ctrlog{}{4} &\ctrlog{}{3} &\ctrlog{}{2} &\ctrlog{}{1} & \gate{S} &\qw   \\
          &\qw  &\qw    &\varbs{1}[\theta_{3}]  &\qw   &\nogate{\qbra{0_\sat}} \qw & \nghost{= \quad} &\qw      &\qw      &\qw      &\targ    &\nogate{\qbra{0_\sat}} \qw  \\
          &\qw  &\varbs{1}[\theta_{2}] &\qw     &\qw   &\nogate{\qbra{0_\sat}} \qw & \nghost{= \quad} &\qw      &\qw      &\targ    &\qw      &\nogate{\qbra{0_\sat}} \qw  \\
          &\varbs{1}[\theta_{1}] &\qw  &\qw     &\qw   &\nogate{\qbra{0_\sat}} \qw & \nghost{= \quad} &\qw      &\targ    &\qw      &\qw      &\nogate{\qbra{0_\sat}} \qw  \\
          &\qw  &\qw                   &\qw     &\qw   &\nogate{\qbra{0_\sat}} \qw & \nghost{= \quad} &\targ    &\qw      &\qw      &\qw      &\nogate{\qbra{0_\sat}} \qw
        }
        }
    \end{split}
\end{equation}
where  hollow circles indicate weight $-1$ CX gates, and the byproduct of stitching and reduction is local, deterministic squeezing $\hat{S}$ on the central mode. Non-zero outcomes result in additional correlated displacements that, combined with the byproduct squeezing, are dealt with in post-processing and decoding (described below).

\emph{MBQC}---At this point, the central modes that comprise the GKP graph state are measured according to the quantum information task at hand. Due to the byproduct squeezing, these raw measurements are performed in transformed bases determined by the desired GKP logical Pauli measurements on $\ket{G}$; as shown in the Supplemental Material, an appropriate homodyne measurement basis can always be found. Once all measurements are performed, the final step is to post-process the raw homodyne outcomes on all modes (satellites and central modes) to obtain canonical outcomes. The post-processing depends on the chosen splitter; the procedure is described for a generic splitter in the End Matter.

\section{Incorporating GKP states on arbitrary lattices}
\label{sec:arbitrary_gkp}

We have so far assumed identical qunaught states~\eqref{eq:dumbbell} at the input to entangling circuits. Targeting GKP states on alternative lattices may result in architectures with fewer noisy or demanding elements or greater tolerance to photon loss~\cite{Noh2019}. Alternatively, one may wish to modify the GKP lattice to intentionally introduce bias that can be leveraged during error correction~\cite{Hanggli2020, Stafford2023, XZZX2021, Zhifei2024}. Here we show how to accommodate for such states within the dumbbell-splitter constructions we have introduced.

Square-lattice GKP states can be transformed via a single-mode Gaussian unitary, $\op{\latG}$, into GKP states on some other lattice. Let us assume the availability of modified qunaught states $\op{\latG}\ket{\varnothing}$. Because identical single-mode Gaussian operations that preserve the origin of phase space commute with a beam splitter, the dumbbell-creation circuit~\eqref{eq:dumbbell} becomes:
\begin{align}
    \label{eq:noisy-dumbells}
    \begin{split}
    \Qcircuit @C=0.3cm @R=0.3cm { 
    \nogate{ \op{\latG} \ket{\qunaught}} & \bsbal{1} & \qw & \qw & \nomultigate{1}{=}&  \nogate{\ket{+}} & \ctrl{1} & \gate{ \latG} & \qw 
    \\
     \nogate{ \op{\latG} \ket{\qunaught}} & \qw & \gate{F} & \qw & \nghost{=}& \nogate{\ket{+}} & \ctrl{-1} & \gate{F \latG  F^\dag} & \qw
    } 
    \end{split} \quad \quad
\end{align}
Notice the asymmetry in the dumbbell: half of the modes have a Fourier transform (type $B$) and half of the modes do not (type $A$). Let us suppose that the type-$A$ (type-$B$) modes meet only with other type-$A$ (type-$B$) modes at macronodes. This is only possible if the target graph $G$ is bipartite, which is the case for foliated CSS codes \cite{Bolt2016}.\footnote{This is not a significant restriction---subsystem codes, which include all stabilizer codes, can be mapped to CSS codes with twice the number of physical and logical qubits, and at least half the code distance~\cite{liu2024}.} Then, the operator $\op{\latG}$ ($\hat F \op{\latG} \hat F^\dag$), which acts identically on all modes of macronodes $A$ ($B$), commutes freely with any network of real beam splitters, including all the splitters we present. Measurement bases can then be chosen such that these lattice-induced operators are undone, reducing the circuit to the square-lattice case.
Therefore, given a set of GKP input states defined uniformly on some non-square lattice, one can construct a measurement pattern that effects the desired measurement on a square-lattice GKP graph state. A complete proof is provided in the Supplemental Material Sec.~\ref{ap:gkp_lattice}. 

\section{Fault-tolerant analysis}
\label{sec:fault_tolerance}

To assess the fault tolerance of the architectures here, let us consider the Gaussian Random Noise (GRN) channel $\mathcal{E}$. This model captures cumulatively the exact effects of anything equivalent to GRN, including loss next to homodyne measurements or amplifiers~\cite{Noh2019,fukui2021all}, while being tractable for large-scale QEC simulations. As an additive Gaussian channel, uncorrelated GRN on each mode can be represented as a covariance matrix $\mat \Sigma_\mathcal{E} = \text{diag}[\epsilon_q, \epsilon_p]$,
where $\epsilon_{q(p)}$ describe added variances in the $\op{q}$ ($\op{p}$) quadratures. We focus on homogeneous noise, \textit{i.e.} $\mathcal{E}$ applied to every qunaught state, capturing state preparation errors.\footnote{A simple (albeit suboptimal) way of handling inhomogeneity is to uniformize the noise by adding Gaussian noise to less noisy modes.}

\emph{Noise propagation}---A uniform GRN channel $\mathcal{E}^{\otimes M}$ spanning all $M$ modes commutes through the dumbbell creation and the splitters to act directly before the homodyne measurements. Uniform losses after state preparation can similarly be commuted and converted into GRN channels by rescaling the measurement outcomes~\cite{fukui2021all, tzitrin2021fault}. Then, noisy homodyne outcomes are obtained by sampling from uncorrelated GRN channels given by 
\begin{align}
\label{eq:updated_covs}
\begin{split}
\mathcal{E}_A &= \op{\latG}^\dagger\mathcal{E} \op{\latG} \\
\mathcal{E}_B &= \hat{F}\op{\latG}^\dagger\mathcal{E}\op{\latG} \hat{F}^\dagger
\end{split}
\end{align}
 in macronodes $A$ ($B$). In the case where $\op{\latG}^\dagger \mathcal{E}\op{\latG}$ is isotropic---for example, square-lattice GKP states with $\epsilon_q = \epsilon_p$---all modes experience the same noise. 

The choice of splitter $\op{U}^{(n)}_{\text{split}}$ has no impact on fault tolerance: splitters that produce the same canonical graph state $\ket{G}$ have the same noise properties (see the End Matter and Supplemental Material Sec.~\ref{ap:noise_propagation} for more details).
For the simplified case of square-lattice GKP states with isotropic noise of strength $\epsilon$, the propagated noise covariance matrix for a macronode is given by $\epsilon(\mat I + \mat A^2)$, where $\mat A$ is the adjacency matrix of a star graph between the central mode and $n-1$ satellite modes.

\emph{Qubit-level errors}---In concatenations with qubit QEC codes~\cite{Xu2023}, we are interested in GKP qubit Pauli error probabilities. These correspond to the likelihood that we have mistakenly obtained the bit 0 rather than 1 (or vice versa) after binning the homodyne outcomes, conditioned on the homodyne values themselves. These quantities are sensitive to the particular GKP error correction scheme being performed, in particular the binning (inner decoding) strategy.
In Ref.~\cite{tzitrin2021fault}, the outcomes from central modes were binned independently to obtain Pauli error probabilities explicitly for the four-splitter; our general form of the propagated noise covariances allows for a more compact calculation, which we present in the Supplemental Material Sec.~\ref{ap:noise_propagation}. In the regime of low $\epsilon$, this decoding strategy results in qubit phase error rates that are bounded approximately by $\operatorname{erfc}(\sqrt{\pi/8n\epsilon}) + n\operatorname{erfc}(\sqrt{\pi/16\epsilon})$, where $n$ is the valence of the node, and $\operatorname{erfc}$ the complementary error function.

\emph{Leveraging quadrature bias}---Expressions~\eqref{eq:updated_covs} for the noise in A- and B-type macronodes point to a mechanism for handling anisotropic noise in the architecture. For uniformly rectangular GKP states with uniform, anisotropic GRN; entanglement generated through splitters; and a target graph state with bipartition into modes $A$ and $B$, we can decouple physical qubit error rates as follows: the phase error rates associated with $X$ measurements on nodes $A$ depend only on the noise along one quadrature of the input states, and those along the nodes $B$ depend on noise along the other quadrature. We present the proof of this claim in the Supplemental Material Sec.~\ref{ap:proof_noise_decouple}. 

There is an immediate consequence of this result for error-corrected MBQC. Foliating a CSS code results in a bipartite graph state, where we may associate partition $A$ ($B$) with primal (dual) modes. Then, for a fault-tolerant memory, error correction can be performed independently on the primal and dual complexes. This is because syndrome extraction requires only $X$ measurements on every node of the graph state, and so only $Z$ errors are relevant in this setting: $X$ errors before $X$ measurements act trivially, and $X$ errors before graph-state creation propagate to $X$ and $Z$ errors before measurements.

Consider a source of GKP states with biased quadrature noise, for example noisier in $\op{q}$ than in $\op{p}$. Associating the primal (dual) qubits with the $\op{q}$ ($\op{p}$) quadratures, we can compensate for this bias through a judicious choice of outer code where we are free to vary the distances corresponding to $X$ ($Z$) errors $d_{X(Z)}$~\cite{Hanggli2020, XZZX2021, Stafford2023, Higgott2023}. For instance, we can tailor the rectangular surface code to the bias in the physical primal and dual errors by modifying its aspect ratio, in this case by setting $d_X > d_Z$.

\begin{figure}[t]
    \centering
{\label{subfig:hyperbolic}\includegraphics[width=.4\textwidth]{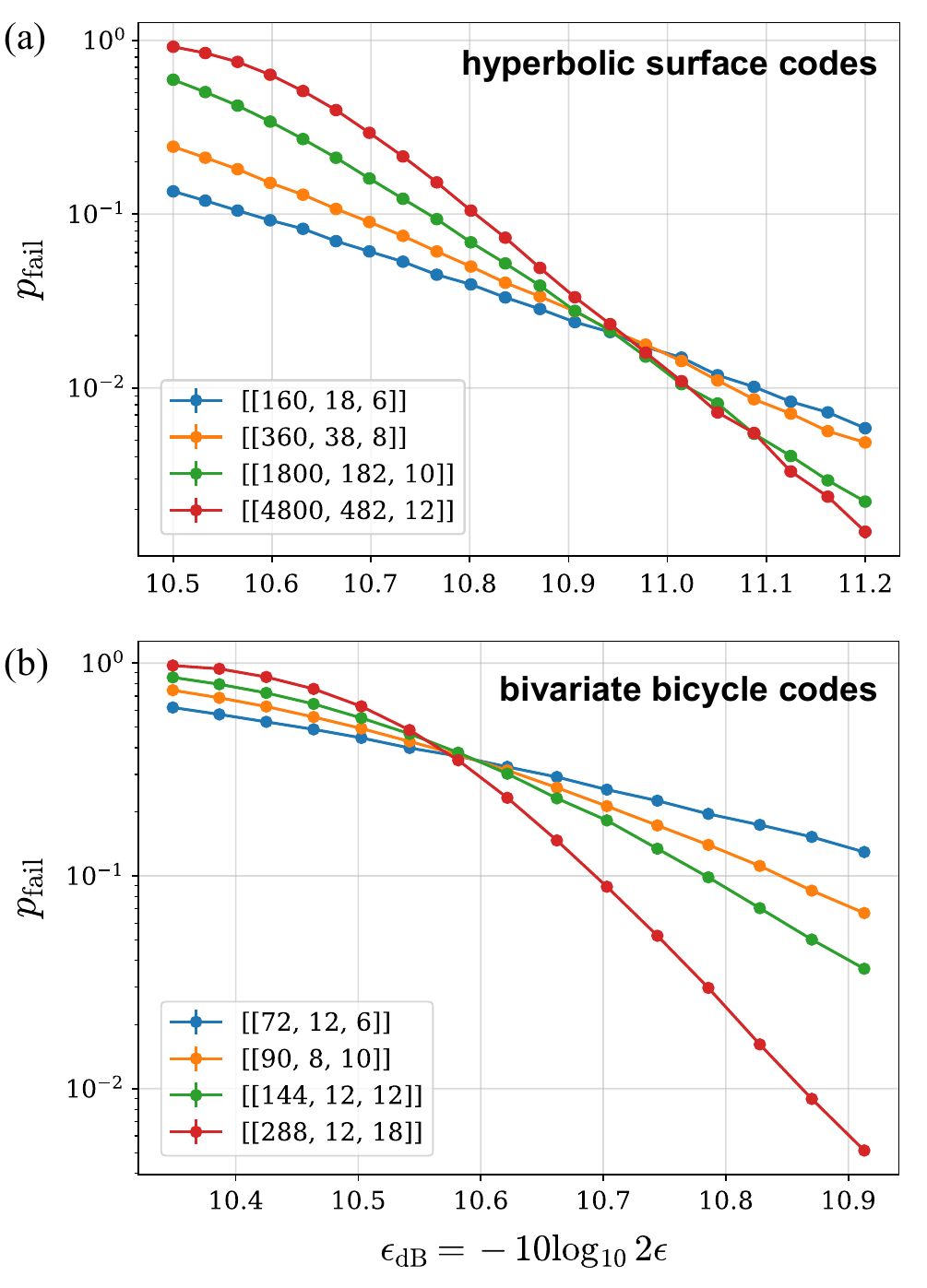}} \qquad
    \caption{Logical error rates, $p_\text{fail}$, for a fault-tolerant memory implemented over (a) a family of hyperbolic surface codes defined on the \{4, 5\} tiling, and (b) bivariate bicycle codes, with uniform GRN on each mode, described by noise parameter $\epsilon$. The foliated graph state is produced using the protocol described in the text with cascade splitters. Each data point comes from 100,000 Monte-Carlo trials with fixed $\epsilon$ and code distance. The simulation was repeated over various even code distances using a correlation-aware inner decoder to translate homodyne outcomes to bit values. $p_\text{err}$ is the fraction of trials in which at least one logical error occurred, and the argument of the logarithm on the $x$-axis is $\epsilon / \epsilon_\text{vac}$ with vacuum variance $\epsilon_\text{vac} = 1/2$. Further details in the Supplemental Material Sec.~\ref{ap:qec_sims}.}
    \label{fig:hyperbolic_sc}
\end{figure}

\emph{Fault-tolerant memory with qLDPC codes}---Having obtained the error probabilities from the inner decoding, we may use them as soft or analog information in an outer (qubit) decoder. Using this two-step decoding scheme we can run error correction simulations for an arbitrary QEC code. For the first code family,  we choose the $\emph{hyperbolic surface codes}$~\cite{freedman2002z2,Breuckmann_2016}, a qLDPC code with favourable encoding rates compared to the 2D surface code~\cite{breuckmann2018phdthesishomologicalquantum}. In particular, we consider the surface code defined over the $\{4, 5\}$ tiling of hyperbolic space, whose foliation results in a graph state with valences 3, 4, and 5. For example, the distance 12 code from this family requires $\sim 19$ physical qubits per logical qubit, compared with 288 physical qubits per logical qubit for the 2D toric code. Our simulations, shown in Fig.~\ref{fig:hyperbolic_sc}, indicate that this code has comparable threshold ($\gtrsim{10.9}$ dB) to the surface code ($\sim \text{10.1 dB}$) in a similar setting~\cite{tzitrin2021fault}. We also consider the \emph{bivariate bicycle codes}~\cite{Bravyi2024},
as these are known to have good thresholds and low overheads, and allow us to apply the dumbbell-splitter constructions to graph state valences of 4, 5, and 6. Indeed, this family of codes has better encoding rates than the hyperbolic surface codes; a distance-12 bivariate bicycle code, for example, requires just twelve physical qubits per logical qubit. Furthermore, despite the increase in valence at certain sites, the QEC threshold is better too, at $\lesssim{10.6}$ dB, making them attractive choices for fault-tolerant designs.
We attribute the slightly higher values to a combination of higher valences in the graph state, resulting in higher qubit-level errors, as well as a higher-valence decoding graph. Details of the simulations are given in the Supplemental Material Sec.~\ref{ap:qec_sims}.

\section{Discussion}

Optical architectures based on GKP qubits enable implementations of arbitrary error-correcting codes using low-depth passive transformations and homodyne measurements. Their key feature is the ability to generate entanglement non-locally, apply Clifford gates, and perform Pauli measurements, all deterministically. This class of architectures is therefore not limited to geometrically local quantum error-correcting codes, opening the door to lower-overhead quantum error correction with qLDPC codes. A prototype of such an architecture, with evidence for its scalability and networkability, can be found in~\cite{Xanadu2024}, where ideas from this work feature in the experiment design.

\section{Acknowledgments}
The authors appreciate the helpful discussions with Eli Bourassa, Austin Daniel, Guillaume Dauphinais, Carlos Gonzalez-Arciniegas, Priya Nadkarni, and Eric Sabo. Computations were performed on the Niagara supercomputer at the SciNet HPC Consortium. SciNet is funded by Innovation, Science and Economic Development Canada; the Digital Research Alliance of Canada; the Ontario Research Fund: Research Excellence; and the University of Toronto. Z. H. was supported by Mitacs through the Mitacs Accelerate program grant. Research at Perimeter Institute is supported in part by the Government of Canada through the Department of Innovation, Science and Economic Development Canada and by the Province of Ontario through the Ministry of
Colleges and Universities.

%\bibliographystyle{IEEEtran}
% \typeout{}
%\bibliography{refs}
% \section*{References}
% \printbibliography[heading=none]

\section{End Matter}

\subsection{Stitching and reduction} 

Let us assume all modes are measured: the satellite modes are measured in $\op{q}$, and the central modes are measured in bases $\op{q}_\theta = \op{R}(\theta) \op{q} \op{R}^\dagger(\theta)$ chosen to implement GKP Pauli measurements according to the MBQC task at hand. This gives a vector of raw homodyne outcomes $\vec{m}$.
In the Supplemental Material Sec.~\ref{ap:notation} -~\ref{ap:app_stitching}, we employ a general LDU decomposition  to show that a splitter-measurement circuit is equivalent to a star-type CV CX network with updated outcomes $\vec m'$ and a different measured quadrature on the central mode, $\op{q}_{\theta'}$. Up to SWAP gates on the inputs that we may ignore, the circuit equivalence is
\begin{align}
\begin{split}\label{eq:stitcherreduction}
     \resizebox*{0.85\columnwidth}{!}{
    \Qcircuit @C=0.3cm @R=0.5cm {
     &  \qw   &\multigate{1}{U_{\text{split}}^{(n)}} & \nogate{\brasub{m_\cent}{q_\theta}}\qw \\
     &{/} \qw &\ghost{U_{\text{split}}^{(n)}}        & \nogate{\qbra{\vec{m}_\sat}} \qw    
    }
    \raisebox{-1.7em}{=} \quad 
    \Qcircuit @C=0.3cm @R=0.5cm {
     &    \qw &\multigate{1}{C_X^\star(\vec{g})} & \nogate{\brasub{m_\cent'}{q_{\theta'}}}\qw \\
     &{/} \qw &\ghost{C_X^\star(\vec{g})} &\nogate{\qbra{\vec{m}'_\sat}} \qw &    
    }
     }
    \end{split},
\end{align}  
where $\op{C}_X^\star (\vec{g}) \coloneqq e^{-i \op{q}_C \otimes \vec{g}^\tp \opvec{p}}$, the central mode acts as the control, the satellite modes are targets, $\opvec{p} = (\op{p}_2,\dots \op{p}_{n})$, and $\vec{g} \in \mathbb{R}^{n-1}$ are the weights.

To connect different macronodes together, each of the $n$ inputs to a splitter receives one half of a dumbbell pair. Using the equivalence in Eq.~\eqref{eq:stitcherreduction} at each macronode, commuting the CZ gates from the dumbbells through the CX networks, and updating the measurement outcomes to $\vec m''$ gives the circuit
  \begin{align}
  \label{eq:fullreduction}
    \begin{split}
    \resizebox{0.8\columnwidth}{!}{
    \Qcircuit @C=0.3cm @R=0.5cm {
     \nogate{\ket{+}^{\otimes n}}   &{/} \qw & \qw &\multigate{1}{C^\star_Z(\vec{h})}      &\nogate{ \brasub{\vec{m}''_{\boldsymbol{N}_C}}{q_{\vec \theta' }} } \qw \\
     \nogate{\ket{+}}    &\qw  & \multigate{1}{C^\star_X(\vec{g})}   &\ghost{C^\star_Z(\vec{h})}  & \nogate{\brasub{m''_\cent}{q_{\theta'}}} \qw \\
     \nogate{\ket{+}^{\otimes n-1}} &{/} \qw & \ghost{C^\star_X(\vec{g})}      & \qw &\nogate{\qbra{\vec{m}''_\sat}} \qw & 
    }
    } \,
    \end{split},
    \end{align}
where $\op{C}^\star_Z(\vec{h}) \coloneqq e^{i \op{q}_\cent \otimes \vec{h}^\tp \opvec{q}_{\vec{N}_\cent}}$ is a star-type CZ network between the central mode of this macronode (middle wire) and the central modes of neighboring macronodes, $\boldsymbol{N}_\cent$ (top wire), with weights $\vec{h}$ arising from the commutation. 
This circuit describes a single stitched node of $G$. For each node in $\boldsymbol{N}_\cent$, there is a $\op{C}^\star_Z$ network coupling it to its neighboring central modes and a $\op{C}^\star_X$ network coupling it to its satellites. Including all modes gives a weighted version of the graph $G$ between the central modes.

Compatibility with GKP encodings requires splitters that yield integer-valued $\vec{g}$, which guarantees integer-valued $\vec{h}$. This ensures both that $C^\star_X$ and $C^\star_Z$ are GKP entangling gates and that $C^\star_X$ acts trivially on the input $\ket{+}$ states.
The Supplemental Material contains constructions for several splitter families that satisfy this property. 

\subsection{Shift matrix} \label{EndMatter:shiftmatrix} 

The canonical outcomes $\vec m''$, which can be regarded as byproduct displacements on the GKP graph state $\ket{G}$, are found by post-processing the vector of raw outcomes, $\vec m$, obtained from the homodyne detectors at the splitters.
These processing rules were calculated explicitly for the four-splitter in Ref.~\cite{tzitrin2021fault}; the technique described here is more general and less cumbersome, lending itself well to simulation and decoder implementation. 

Consider a macronization with $N$ total modes (satellite and central), and assume that that the central modes are measured in $\op{p}$ (logical $X$), leaving the more general case to the Supplemental Material Sec.~\ref{ap:app_stitching}. First, populate a length $2N$ vector $\tilde{\vec{m}}= [\vec m_q, \vec m_p]^\tp$ with the $N$ measurement outcomes such that the $i$-th entry of $\vec m_q$ ($\vec m_p$) is only non-zero if the $i$-th mode is measured in position (momentum). The length $2N$ vector of canonical outcomes $\tilde{\vec{m}}''$ (in which $\vec{m}''$ is embedded in the same way) is given by
\begin{align}\label{eq:total_outcomes}
    \tilde{\vec m}'' 
    = 
    \underbrace{\mat S_{G} \mat S_{\text{CX}} \mat S_{\text{DB}}^{-1} \mat S_\text{split}^{-1}}_{\mat{S}_\text{shift}} \tilde{\vec m}.
\end{align}
From left to right, the $\mat{S}_\alpha \in \mathbb{R}^{2n\times 2n}$ are the symplectic matrices associated with the CZ network of the graph $G$ between the central nodes; the CX networks between central nodes and their satellites; the CZ network of the input dumbbells [Eq.~\eqref{eq:dumbbell}]; and finally $\op{U}_\text{split}$, the splitters associated with every macronode. We identify the composite matrix $\mat{S}_\text{shift}$ from Eq.~\eqref{eq:total_outcomes} as the \textit{shift matrix}, which stores information useful for inner decoding---obtaining qubit values from homodyne outcomes. In the noiseless case with ideal GKP states, $\mat{S}_\text{shift} \vec{\tilde m}$ contains integer multiples of $\sqrt{\pi}$. The bit value corresponding to a Pauli $X$ measurement on the canonical mode with index $k$ is the parity of $\left(\mat{S}_\text{shift} \vec{\tilde m}\right)_{k+N}$ modulo $\sqrt{\pi}$, where $N$ is the total number of modes. 

\subsection{Equivalence of fault tolerance}
For our noise model, varying the splitter $\op{U}^{(n)}_{\text{split}}$ has no impact on fault tolerance, provided it reduces to the same CX network. This follows because the reduction is state independent, so the noise can be propagated through $\hat C^\star_X(\vec{g})$ in circuit~\eqref{eq:stitcherreduction} and applied directly to the measurement outcomes $\vec{m}_S'$ and $m_C'$. Then, for integer-valued $\vec{g}$, the entangling circuit consists entirely of GKP qubit gates, and the outcomes can be corrected to the nearest integer multiple of $\sqrt{\pi}$. In general, the GRN channel for a macronode of size $n$ and of type $A$ and $B$ are given by $\mathcal{E}'_A = \hat{C}_X^\star(\vec{g})(\op{\latG}^\dagger\mathcal{E} \op{\latG})^{\otimes n} \hat{C}_X^{\star \dagger}(\vec{g})$ and $
\mathcal{E}'_B = \hat{C}_X^\star(\vec{g})(\hat{F} \op{\latG}^\dagger\mathcal{E} \op{\latG} \hat{F}^\dagger)^{\otimes n}\hat{C}_X^{\star \dagger}(\vec{g})$.

%%%%%%%%%% Merge with supplemental materials %%%%%%%%%%

\onecolumngrid
\newpage
\begin{center}
\vspace{2em}
\textbf{\large Supplemental Material for: Linear-optical quantum computation with arbitrary error-correcting codes}
\vspace{4em}
\end{center}
\twocolumngrid
%%%%%%%%%% Merge with supplemental materials %%%%%%%%%%
%%%%%%%%%% Prefix a "S" to all equations, figures, tables and reset the counter %%%%%%%%%%
\setcounter{equation}{0}
\setcounter{figure}{0}
\setcounter{table}{0}
\setcounter{section}{0}
\setcounter{page}{1}
\renewcommand{\thefigure}{S\arabic{figure}}
\renewcommand{\theHfigure}{S\arabic{figure}}
\counterwithin*{equation}{section}
\renewcommand{\theequation}{S\Roman{section}\arabic{equation}}
%%%%%%%%%% Prefix a "S" to all equations, figures, tables and reset the counter %%%%%%%%%%

\section{Notation, conventions, and Gaussian operations}\label{ap:notation}

Position and momentum operators are defined in terms of the raising and lowering operators as $\op q = \tfrac{1}{\sqrt{2}}(\op a + \op a^{\dag})$ and $\op p = \tfrac{-i}{\sqrt{2}}(\op a - \op a^{\dag})$; here $[\op q, \op p]=i$, and $\hbar=1$. The eigenstates are labeled $\qket{s}$ and $\pket{t}$, so that $\op q \qket{s}=s\qket{s}$ and $\op p \pket{t}=t\pket{t}$.

The Heisenberg action of an $n$-mode Gaussian unitary operator, $\op U$, on an $n$-mode vector of quadrature operators $\opvec x \coloneqq (\opvec{q}, \opvec{p})^\tp $ with $ \opvec{q} \coloneqq (\op{q}_1, \op{q}_2, \dots \op{q}_n)^\tp $ and $ \opvec{p} \coloneqq (\op{p}_1, \op{p}_2, \dots \op{p}_n)^\tp$ is given by the linear transformation 
\begin{equation} \label{symplectic}
\begin{aligned}
	\op U^{\dag} \opvec x \op U \mapsto \mat{S}_{U} \opvec x
\end{aligned}
\end{equation}
where $\mat{S}_{U} \in \mathbb{R}^{2n \times 2n}$ is a symplectic matrix. The symplectic matrix for successive Gaussian unitaries is found by composing the matrices in Schrodinger-picture order, \emph{e.g.} $\op{U}_2 \op{U}_1 \mapsto \mat{S}_2 \mat{S}_1$. 

Similarly, a completely-positive Gaussian map transforms the means $\vec{r}$ and the covariance matrix $\mat \Sigma$ of a Gaussian state  through~\cite{serafini2017quantum}:
\begin{align}
\vec{r} &\mapsto \mat X \boldsymbol{r} + \vec{d}
\\
\mat \Sigma &\mapsto \mat X \mat \Sigma \mat X^\tp + \mat Y \label{appeq:Gaussianchannel}
\end{align}
where $\mat{X}$ and $\mat{Y}$ are matrices, and $\vec{d}$ is a vector of displacements. For a Gaussian random noise channel $\mathcal{E}$ with covariance matrix $\mat \Sigma_\mathcal{E}$, we have
    \begin{align}
    \mathcal{E} \to \begin{cases} \mat X=\mat I, \vec{d} = \vec{0} \\ \mat Y= \mat \Sigma_\mathcal{E}
    \end{cases}.
    \end{align}
Similarly, for a Gaussian unitary $\op{U}$:
    \begin{align}
    \op{U} \to \begin{cases} \mat X= \mat{S}_U \\ \mat Y = \mat 0, \end{cases}
    \end{align}
where $\mat S_U$ is the symplectic matrix from Eq.~\eqref{symplectic} associated with $\op{U}$. In circuit diagrams, we use the shorthand notation $ \op{B} \mathcal{E} \op{A} \rightarrow \op{B} \mathcal{E} (\op{A} ( \cdot ) \op{A}^\dagger ) \op{B}^\dagger$ when composing channels and unitaries.

A displacement operator over $n$ modes is defined as
 \begin{align} \label{multimodedisplacement_alt}
		\op{D}(\mathbf{z}) 
            & \coloneqq \exp \big[ i\sqrt{2}( \vec{\alpha}_I^\tp \opvec{q} - \vec{\alpha}_R^\tp \opvec{p} ) \big] , 
	\end{align}
with the displacement amplitudes characterized by a 2$n$-dimensional vector $\mathbf{z} \coloneqq (\vec{\alpha}_R \, \, \vec{\alpha}_I )^\tp$, where $\vec{\alpha}_R$ and $\vec{\alpha}_I$ are each $n$-dimensional column vectors of real numbers. 
Under a Heisenberg-picture transformation by a Gaussian unitary $\op{U}$, the amplitudes are transformed by the inverse of the symplectic matrix,
    \begin{align} \label{eq:displacementsymplecticrule}
        \op{U}^\dagger \op{D}(\mathbf{z}) \op{U} = \op{D}(\mat{S}_{U}^{-1} \mathbf{z}) .
    \end{align}

The squeezing operator
\begin{align}\label{squeezeOp}
    \op S(\zeta) \coloneqq e^{-\tfrac{i}{2}(\ln\zeta)(\op q \op p + \op p \op q)},
\end{align}
acts on eigenstates as $\hat{S}(\zeta) \qket{s} = \qket{\zeta s}$ and $\hat{S}(\zeta) \pket{t} = \pket{\zeta^{-1} t}$~\cite{Walshe2020}. Its Heisenberg action is $\op S^{\dag}(\zeta)\op q \op S(\zeta) = \zeta \op q$ and $\op S^{\dag}(\zeta)\op p \op S(\zeta) = \tfrac{1}{\zeta} \op p$.

A phase delay by $\theta$ (also commonly referred to as a rotation) is produced by the operator
	\begin{align} \label{phasedelay}
		\op{R}(\theta) \coloneqq e^{i \theta \op{a}^\dagger \op{a}} =  e^{i \frac{\theta}{2} (\op{q}^2 + \op{p}^2 - 1)}
	\end{align}
The Heisenberg action on the position quadrature gives the rotated quadrature,
	\begin{align} \label{eq:rotquadop}
		\op{q}_\theta \coloneqq \op{R}^\dagger(\theta) \op{q} \op{R}(\theta) =& \cos \theta \op{q} -  \sin \theta \op{p}
	 \end{align}	
with phase-advanced position eigenstates,
    \begin{align}\label{eq:phasedelaykets}
		\ket{s}_{q_\theta} \coloneqq \op{R}^\dagger(\theta) \ket{s}_q 
	\end{align}
satisfying $\op{q}_\theta \ket{s}_{q_\theta} = s \ket{s}_{q_\theta}$. A special case is the 
Fourier transform operator
\begin{align}
    \op F \coloneqq \op{R}(\tfrac{\pi}{2}) = e^{i \frac{\pi}{2} \hat{n} } 
\end{align}
that rotates the position and momentum quadratures by $\tfrac{\pi}{2}$: $\op F^\dag \op q \op F=-\op p$ and $\op F^\dag \op p \op F=\op q$.

A CV controlled-X (CX) gate between modes $j$ and $k$ with weight $g$ is 
\begin{align} \label{eq:CXgate}
    \CX^{jk}(g) \coloneqq e^{-i g \op q_j \otimes \op p_k} 
    =  \quad 
    \begin{split}
        \Qcircuit @C=0.4cm @R=0.7cm
        {
         & \ctrlg{g}{1} & \qw \quad j \\
         & \targ        & \qw \quad k
         }
         \end{split}\quad .
\end{align}
The weight of a CX gate can be changed using local squeezing,
	\begin{align}
		\op{S}_j^\dagger(\zeta) \CX^{jk}(g) \op{S}_j(\zeta) = \op{S}_k(\zeta) \CX^{jk}(g) \op{S}^\dagger_k(\zeta) = \CX^{jk}(\zeta g)
	\end{align}
described by the circuit identities,
\begin{equation}\label{eq:squeezingcontrolledgate}
\begin{split}
\resizebox{0.8\columnwidth}{!}{
\Qcircuit @C=0.6em @R=1.5em {
		&\gate{S(\zeta)}&\ctrlg{g}{1}& \gate{S^\dagger(\zeta)}& \qw && \raisebox{-3em}{=}&&&\qw&\ctrlg{g}{1}& \qw& \qw &&\raisebox{-3em}{=}&& &\ctrlg{\zeta g}{1}& \qw\\
		&\qw& \targ&\qw &\qw&&&&&\gate{S^\dagger(\zeta)}& \targ&\gate{S(\zeta)} &\qw&&&&& \targ{-1}&\qw\\
	}
  }
\end{split}
\; .
\end{equation}
Commuting controlled-X gates through one another can introduce a third gate,
\begin{align}
    \CX^{23}(a) \CX^{12}(b) = \CX^{12}(b) \CX^{23}(a) \CX^{13}(-ab) \label{eq:commuteCX_op}
\end{align}
described by the circuit
\begin{align}\label{commuteCX}
\begin{split}
    \Qcircuit @C=0.5cm @R=0.5cm
    {
     & \qw          & \ctrlg{b}{1} & \qw                                           & \qw & \\
     & \ctrlg{a}{1} & \targ        & \qw                                           & \qw & \\
     & \targ        & \qw          & \qw                                           & \qw &
    }
    \raisebox{-2em}{=} \quad
    \Qcircuit @C=0.5cm @R=0.5cm
    {
     & \ctrlg{b}{1} & \qw          & \ctrlg{-ab}{2}[0.25] & \qw & \\
     & \targ        & \ctrlg{a}{1} & \qw                                           & \qw & \\
     & \qw          & \targ        & \targ                                         & \qw &
    }
    \end{split}.
\end{align}

A CV controlled-Z (CZ) gate with weight $g$ is
\begin{align} \label{eq:CZgate}
    \CZ^{jk}(g) \coloneqq e^{i g \op q_j \otimes \op q_k} 
    =  \quad 
    \begin{split}
        \Qcircuit @C=0.4cm @R=0.7cm
        {
         & \ctrlg{g}{1} & \qw \quad j \\
         & \ctrl{-1}        & \qw \quad k
         }
         \end{split}\quad .
\end{align}
A network of CZ gates over many modes is given by $\op C_Z(\mat A) = e^{\frac{i}{2} \vec{\op q}^\tp \mat A \vec{\op q}}$ where $\mat A$ is an adjacency matrix found in the lower left block of the associated symplectic matrix.\footnote{This notation differs from that in the main text, where the star-type CZ network takes a vector of weights $\vec{g}$ as input.}
An identity significant to our presentation, particularly Eq.~(A2) from the main text and Eq.~\eqref{circ:canon_graph_meas} herein, is the commutation relation between a CX gate and a CZ gate with generic weights,
\begin{align}\label{commuteCXCZ}
\begin{split}
    \Qcircuit @C=0.5cm @R=0.5cm
    {
     & \qw          & \ctrlg{b}{1} & \qw                                           & \qw & \\
     & \ctrlg{a}{1} & \targ        & \qw                                           & \qw & \\
     & \ctrl{0}        & \qw          & \qw                                           & \qw &
    }
    \raisebox{-2em}{=} \quad
    \Qcircuit @C=0.5cm @R=0.5cm
    {
     & \ctrlg{b}{1} & \qw          & \ctrlg{-ab}{2}[0.25] & \qw & \\
     & \targ        & \ctrlg{a}{1} & \qw                                           & \qw & \\
     & \qw          & \ctrl{0}        & \ctrl{0}                                         & \qw &
    }
    \end{split}.
\end{align}

A real beam splitter is defined as
\begin{align}\label{appeq:beamsplitter}
\op{B}_{jk}(\theta)\coloneqq e^{-i\theta(\op q_j \op p_k - \op p_j \op q_k)} = \quad 
\begin{split} 
    \Qcircuit @C=.7cm @R=1cm
    {
     & \varbs{1} & \qw \quad j  \\
     & \qw       & \qw  \quad k
    }
\end{split}
\end{align}
with \emph{transmission angle} $\theta$. For balanced (50:50) beam splitters ($\theta = \pi/4$), the arrow in the circuit is unlabeled.

\section{Gate identities via LDU-type decompositions}
\label{ap:LDU}

Performing LDU and UDL decompositions on the matrix blocks of the symplectic matrix for a single beam splitter gives the useful gate identities~\cite{Walshe2021}:
\begin{subequations} \label{beamsplitterLDU}
\begin{align} 
    \op{B}_{jk}(\theta) 
        & = \CX^{jk}(t_\theta) \big[ \op{S}_j^\dagger(s_\theta) \otimes \op{S}_k(s_\theta) \big] \CX^{kj}(-t_\theta) \\
        & = \CX^{kj}(-t_\theta) \big[ \op{S}_j(s_\theta) \otimes \op{S}^\dagger_k(s_\theta) \big] \CX^{jk}(t_\theta)
    \, ,
\end{align}
\end{subequations}
with $t_\theta \coloneqq \tan \theta$ and $s_\theta \coloneqq \sec \theta$, also described as circuits
\begin{align}\label{BSdecomp}
    \begin{split}
        \Qcircuit @C=0.4cm @R=1cm
        {
         & \varbs{1} & \qw   \\
         & \qw       & \qw 
        }
        &
        \quad \raisebox{-1.5em}{=} \quad
        \Qcircuit @C=0.5cm @R=0.5cm
        {
        &\ctrlg{\tan{\theta}}{1} &\gate{S(\sec{\theta})}      &\targ                     &\qw \\
        &\targ                   &\gate{S^\dag(\sec{\theta})} &\ctrlg{-\tan{\theta}}{-1} &\qw
        }
    \end{split} 
    \\
    \begin{split}
        &\raisebox{-1.5em}{=} \quad
        \Qcircuit @C=0.5cm @R=0.5cm
        {
        &\targ                    &\gate{S^\dag(\sec{\theta})} &\ctrlg{\tan{\theta}}{1} &\qw \\
        &\ctrlg{-\tan{\theta}}{-1} &\gate{S(\sec{\theta})}     &\targ                   &\qw
        }
    \end{split}.
\end{align}
From here, we see that measuring one mode of a beam splitter in position induces a CX gate along with squeezing and a correlated shift on the other mode:
\begin{align}\label{eq:BSdecomp_qmeas}
    \begin{split}
    \resizebox{0.85\columnwidth}{!}{
           \Qcircuit @C=0.5cm @R=1cm
         {
        & \varbs{1} & \qw \\
        & \qw       & \rstick{\qbra{m}} \qw 
        }
        \quad \quad \raisebox{-1.3em}{=} \;
         \Qcircuit @C=0.4cm @R=0.5cm
        {
         &\ctrlg{\tan{\theta}}{1} &\gate{S(\sec{\theta})}              &\gate{X^\dagger(m\tan \theta) }\qw &\qw            \\
         &\targ                   &\rstick{\qbra{m \cos{\theta}}}  \qw 
        }
        }
    \end{split},
\end{align}
where $\op{X}(s) = e^{-is \op{p}}$ is a position-shift operator.

Here, we generalize this procedure to give gate decompositions for certain Gaussian unitaries over many modes. 
In the basis $(\hat{q}_1, \ldots, \hat{q}_n, \hat{p}_1 \ldots \op{p}_n)$, a Gaussian unitary $\op{U}$ over $n$ modes that does not mix position and momentum has a symplectic matrix of the form
    \begin{align} \label{appeq:symplecticmatrices}
        \mat{S}
        = \begin{bmatrix}   \mathbf M & \mathbf{0} \\ \mathbf{0} & \mathbf{M}^{-\tp}  \end{bmatrix},
    \end{align}
where $\mat{M} \in \mathbb{R}^{n \times n}$.
Examples include the parity operator, CX gates~\eqref{eq:CXgate}, beam splitters of the form in~\eqref{appeq:beamsplitter}, and any combination of the above over any number of modes. Passive Gaussian unitaries (such as beam splitter networks) have the additional property that $\mat{M}^{-1} = \mat{M}^\tp$, so $\mat{S} = \mat{M} \oplus \mat{M} = \mat{I}_2 \otimes \mat{M}$. 

Consider a two-mode CX gate $\hat{C}_X^{jk}(g)$. When the mode ordering is such that $j<k$, the block matrix in Eq.~\eqref{appeq:symplecticmatrices} is lower-triangular,
    \begin{align} \label{appeq:MCX}
        \mat{M}_{\text{CX}\downarrow} = \begin{bmatrix}   1 & 0 \\ g & 1  \end{bmatrix},
    \end{align}
where the arrow $\downarrow$ is an indicator of the mode-label ordering.
When $j>k$, the block matrix is upper triangular $\mat{M}_{\text{CX}\downarrow} = \mat{M}_{\text{CX}\uparrow}^\tp$. More generally, we consider CX networks over many modes. If each gate in the network satisfies $j<k$, then the block matrix for the entire network will be described by a lower triangular matrix.
Similarly, if all the gates satisfy $j>k$, the block matrix will be upper triangular. However, if a network has CX gates of both types, the block matrix will be neither lower nor upper triangular.

Any square matrix $\mat{M}$ can be factorized as $\mat{M} = \mat{P} \mat{L} \mat{D} \mat{U}$, where $\mat{P}$ is a permutation matrix, $\mat{L}$ is lower triangular, $\mat{D}$ is diagonal, and $\mat{U}$ is upper triangular. Inserting this decomposition into Eq.~\eqref{appeq:symplecticmatrices} gives
    \begin{align}
\mathbf S 
=
\begin{bmatrix} \mathbf P & \mathbf 0 \\ \mathbf 0 & \mathbf P \end{bmatrix}
\begin{bmatrix} \mathbf L & \mathbf 0 \\ \mathbf 0 & \mathbf L^{-\tp} \end{bmatrix}
\begin{bmatrix} \mathbf D & \mathbf 0 \\ \mathbf 0 & \mathbf D^{-1} \end{bmatrix}
\begin{bmatrix} \mathbf U & \mathbf 0 \\ \mathbf 0 & \mathbf U^{-\tp} \end{bmatrix},
    \end{align}
noting that $\mat{P}^{\tp} = \mat{P}^{-1}$, and $\mat{D}^\tp = \mat{D}$.
The Hilbert-space interpretation is a gate decomposition of $\op{U}$ into four Gaussian unitaries:
    \begin{align} \label{appeq:LDUdecomp}
        \op{U} = \op{U}_\text{SWAP} \op{U}_{\text{CX} \downarrow} \op{U}_\text{sq} \op{U}_{\text{CX} \uparrow},
    \end{align}
where $\op{U}_{\text{SWAP}}$ describes a network of CV swap gates, $\op{U}_{\text{CX} \downarrow}$ describes a network of CX gates with $j < k$ for each gate, and $\op{U}_{\text{CX} \uparrow}$ is a different CX network with $j < k$ for each gate. Unlike in a CZ network, the individual CX gates in $\op{U}_{\text{CX} \downarrow}$ (and $\op{U}_{\text{CX} \uparrow}$) may not commute---specifically, this occurs when the control mode of one gate is the same as the target mode of a different gate, as in Eq.~\eqref{commuteCX}. The middle unitary describes single-mode squeezing on each mode $\op{U}_\text{sq} = \bigotimes_{k=1}^n \op{S}_k(\zeta_k)$. 

Similarly, we can use the UDL decomposition, $\mat{M} =  \mat{U}' \mat{D}' \mat{L}' \mat{P}'$ to find a different decomposition $\mat{U}$, where the roles of the CX networks are swapped:
  \begin{align} \label{appeq:UDLdecomp}
        \op{U} =  \op{U}'_{\text{CX} \uparrow} \op{U}'_\text{sq} \op{U}'_{\text{CX} \downarrow} \op{U}'_\text{SWAP}.
    \end{align}
The unitaries in this decomposition may not be the same as those in Eq.~\eqref{appeq:UDLdecomp}, \textit{e.g.} $\op{U}'_{\text{CX} \downarrow} \neq \op{U}_{\text{CX} \downarrow}$. 
However, for real beamsplitter networks $\mat{M}^{-1} = \mat{M}^\tp$, which gives $\mat{L}' = \mat{U}^{-\tp}$, $\mat{U}' = \mat{L}^{-\tp}$, $\mat{D}' = \mat{D}^{-1}$, and $\mat{P}' = \mat{P}$.
The relations between LDU and UDL decompositions of a single beam splitter are exemplified in Eqs.~\eqref{beamsplitterLDU}.

\section{Stitching and reduction}
\label{ap:app_stitching}

We show here how splitters and position measurements are used to reduce a GKP graph state to another GKP graph state over fewer modes with modified connectivity. 

\subsection{Measuring the splitter} 

We use the UDL decomposition~\eqref{appeq:UDLdecomp} to express the splitter in terms of four Gaussian unitaries,
$\op{U}_\text{split}^{(n)} = \op{U}_{\text{CX} \uparrow} \op{U}_\text{sq} \op{U}_{\text{CX} \downarrow} \op{U}_{\text{SWAP}}$.
The first network of CX gates distributes shifts according to the measurement outcomes on the satellite modes. This is because all CX gates in $\op{U}_{\text{CX} \uparrow}$ have a position-measured control mode, $\brasub{m_\cent}{q_\theta} \otimes \qbra{\vec{m}_\sat} \op{U}_{\text{CX} \uparrow} = \brasub{m'_\cent}{q_\theta} \otimes \qbra{\vec{m}_\sat'}$ where $m_\cent'$ and  $\vec{m}_\sat'$ are linear combinations of $m_\cent$ and  $\vec{m}_\sat$. 

The squeezings from $\op{U}_\text{sq}$ on the satellite modes rescale their measurement according to $\qbra{m_\sat} \hat{S}(\zeta) = \qbra{m_\sat/\zeta}$. The squeezing on the central mode, which we call the \emph{terminal squeezing}, both rescales the outcome and shears the measurement basis, since the squeezing is not in general aligned with the principal axis of $\op{q}_\theta$. For this reason, we leave the terminal squeezing $\op{S}(\zeta)$ in the circuit and address its effect later. 

Finally, we use the commutation relation in Eq.~\eqref{commuteCX} to push all CX gates in $\op{U}_{\text{CX} \downarrow}$ whose control-mode is a satellite mode onto the measurements.
The commutation relations create new CX gates; we gather these gates into a star-type CX network
$\CX^\star (\vec{g}) \coloneqq e^{-i \op{q}_\cent \otimes \vec{g}^\tp \opvec{p}}$, with the central mode as the control and the satellite modes as targets, $\vec{g} \in \mathbb{R}^{n-1}$ is a vector of weights, and $\opvec{p} = (\op{p}_2,\dots \op{p}_{n})$.
Pushing all remaining shifts on the central mode through the terminal squeezing and updating the outcomes on all modes ($\vec{m}_\sat \rightarrow \vec{m}_\sat'$ and $m_C \rightarrow m_\cent^{(I)}$), we get the circuit equivalence
\begin{align}
\begin{split}\label{appeq:stitcherreduction}
    \resizebox*{0.85\columnwidth}{!}{
    \Qcircuit @C=0.3cm @R=0.5cm {
     &  \qw   &\multigate{1}{U_{\text{split}}^{(n)}} & \nogate{ \brasub{m_\cent}{q_\theta}} \qw \\
     &{/} \qw &\ghost{U_{\text{split}}^{(n)}}             & \nogate{\qbra{\vec{m}_\sat}} \qw    
    }
    \raisebox{-1.8em}{=} \; 
    \Qcircuit @C=0.3cm @R=0.5cm {
     &    \qw &\multigate{1}{U_{\text{SWAP}}} &\multigate{1}{C_\text{X}^\star(\vec{g})} &\gate{S(\zeta)}         &\nogate{ \brasub{m_\cent^{(I)}}{q_\theta} } \qw \\
     &{/} \qw &\ghost{U_{\text{SWAP}}}        &\ghost{C_\text{X}^\star(\vec{g})}        &\qw &\nogate{\qbra{\vec{m}_\sat'}} \qw     
    }
    }
    \end{split},
\end{align}

The above circuit shows that a physical measurement of $\op{q}_\theta$ on the central mode (left-hand side) corresponds to a measurement of the transformed quadrature $\op{S}^\dagger(\zeta) \op{q}_\theta \op{S}(\zeta) $ on the reduced circuit (right-hand side). 
We use Eq.~\eqref{eq:simmeasurement} to absorb the terminal squeezing into the central-mode measurement, $\brasub{m_\cent^{(I)}}{q_\theta} \op{S}(\zeta) = \brasub{m_\cent''}{q_{\theta'}} $, which changes the measured quadrature and modifies the outcome ($\op{q}_\theta \rightarrow \op{q}_{\theta'}$ and $m_\cent^{(I)}\rightarrow m_C'$):
\begin{align}
\begin{split}\label{appeq:stitcherreduction2}
    \resizebox*{0.85\columnwidth}{!}{
    \Qcircuit @C=0.3cm @R=0.5cm {
     &  \qw   &\multigate{1}{U_{\text{split}}^{(n)}} & \nogate{ \brasub{m_\cent}{q_\theta}} \qw \\
     &{/} \qw &\ghost{U_{\text{split}}^{(n)}}             & \nogate{\qbra{\vec{m}_\sat}} \qw    
    }
    \raisebox{-1.8em}{=} \;
    \Qcircuit @C=0.3cm @R=0.5cm {
     &    \qw &\multigate{1}{U_{\text{SWAP}}} &\multigate{1}{C_\text{X}^\star(\vec{g})} &\nogate{\brasub{m'_\cent}{q_{\theta'}} } \qw \\
     &{/} \qw &\ghost{U_{\text{SWAP}}}        &\ghost{C_\text{X}^\star(\vec{g})}        &\nogate{\qbra{\vec{m}_\sat'}} \qw     
    }
    }
    \end{split}.
\end{align}  
The specific measured quadrature on the central mode in the physical circuit will depend on the desired canonical GKP qubit measurement there; see Sec.~\ref{appsec:logmeas}. Here we can leave it unspecified. 

Notice that the circuit~\eqref{appeq:stitcherreduction2} is agnostic to the input states, relying only on properties of the splitter and the measurements. The SWAP gates contained in $\hat U_\text{swap}$ can be treated several ways: (1) propagate them through the CX network and permute the measurements, or (2) propagate them to the input states and permute those instead. We opt for the latter in our analysis because we work with a set of uniformly defined input states, which are permutation-invariant. In cases where the input states vary and their precise placement matters (for example, if some GKP states are replaced by momentum-squeezed states~\cite{Tzitrin2020}), these SWAPs must be accounted for in the routing stage.

\subsection{Obtaining the target GKP graph state}

Let us now include the input states in our discussion. Consider a collection of $2n$ modes prepared in GKP $\ket{+}$ states entangled via a network of CZ gates of arbitrary weight and connectivity.
This prepares a graph state $\CZ(\mat{A}) \ket{+}^{\otimes 2n}$, which is a square-lattice GKP graph state when $\mat{A}$ is integer valued. A special case of this is a collection of $n$ GKP dumbbells, mutually separable, each taking the form $\CZ(1)\ket{+}\otimes\ket{+}$, as in Eq.~(1) from the main text.
Sending $n$ of these modes to a splitter is described by the circuit
    \begin{align}
    \begin{split} \label{eq:CZnetworkstitch}
    \resizebox{0.8\columnwidth}{!}{
    \Qcircuit @C=0.3cm @R=0.65cm {
     \nogate{\ket{+}^{\otimes n}}   &{/} \qw &\multigate{2}{C_\text{Z}(\mat{A})}      &\rstick{\text{(neighbors)}} \qw                         & \\
     \nogate{\ket{+}}                &\qw     &\ghost{\op{C}_\text{Z}(\mat{A})}       &\multigate{1}{U_\text{split}^{(n)}} &\nogate{ \brasub{m_\cent}{q_{\theta}} }\qw                         &  \\
     \nogate{\ket{+}^{\otimes n-1}} &{/} \qw &\ghost{C_\text{Z}(\mat{A})}        &\ghost{U_\text{split}^{(n)}}        &\nogate{\qbra{\vec{m}_\sat}} \qw &
    }
    }
    \end{split},
    \end{align}
where the other $n$ modes (top wire) are potentially sent to neighboring splitters before being measured in the appropriate basis, and the middle wire in the circuit is the central mode of the given macronode.
Using the circuit equivalence~\eqref{appeq:stitcherreduction2}, we obtain
      \begin{align}
        \begin{split}
        \resizebox{0.8\columnwidth}{!}{
    \Qcircuit @C=0.3cm @R=0.65cm {
      \nogate{\ket{+}^{\otimes n}}   &{/} \qw &\multigate{2}{C_\text{Z}(\mat{A}')}
      &\rstick{ \text{(neighbors)} } \qw             & \\
      \nogate{\ket{+}} &\qw     &\ghost{\op{C}_\text{Z}(\mat{A}')}        &\multigate{1}{C_\text{X}^\star(\vec{g})} &\nogate{ \brasub{m'_\cent}{q_{\theta'}}}\qw &   \\
      \nogate{\ket{+}^{\otimes n-1}} &{/} \qw &\ghost{C_\text{Z}(\mat{A}')}        &\ghost{C_\text{X}^\star(\vec{g})}               &\nogate{\qbra{\vec{m}_\sat'}} \qw              & 
    }
    }
        \end{split}\,,
    \end{align}
where (per the discussion in the previous subsection) we have applied the SWAP gates to the input states, leaving them invariant, and $\mat{A}' = (\mat{I}_n \oplus \mat{P}) \mat{A} (\mat{I}_n \oplus \mat{P})$, where $\mat{P}$ is the permutation matrix associated with the SWAP network (the upper block of the block-diagonal SWAP network).

Conjugating $\CZ(\mat{A}')$ with the splitter's CX network, $\CX^\star(\vec{g})$, gives a new CZ network over $n$ modes with adjacency matrix $\mat{A}'' = (\mat{I}_n \oplus \mat{M}_{\text{CX}^\star}^\tp) \mat{A}' (\mat{I}_n \oplus \mat{M}_{\text{CX}^\star})$, where $\mat{M}_{\text{CX}^\star}$ is the upper block of the block-diagonal symplectic matrix for $\op C_X^\star$ (for a single CZ gate, refer to the identity in Eq.~\eqref{commuteCXCZ}). The position measurements in the macronode delete most of these CZ gates, propagating shifts to the neighboring macronodes. If this procedure is performed at every macronode, the remaining gates at each macronode comprise a CZ network $\CZ^\star(\vec{h})$, with weights $\vec{h}$. This corresponds to a star graph across central mode $\cent$ and $\vec N_\cent$, the central modes of neighboring macronodes:
    \begin{align}
    \label{circ:canon_graph_meas}
    \begin{split}
    \resizebox{0.85\columnwidth}{!}{
    \Qcircuit @C=0.3cm @R=0.65cm {
      \nogate{\ket{+}^{\otimes n}}   &{/} \qw &\qw                                      &\multigate{1}{C^\star_Z(\vec{h})}                               &\nogate{\brasub{\vec{m}_{\vec{N}_\cent}''}{ q_{\vec{\theta'}} }}\qw                         &\\
      \nogate{\ket{+}} &\qw     &\multigate{1}{C_\text{X}^\star(\vec{g})} &\ghost{C^\star_Z(\vec{h})}  & \nogate{ \brasub{m''_\cent}{q_{\theta'}} }\qw          &   \\
      \nogate{\ket{+}^{\otimes n-1}} &{/} \qw &\ghost{C_\text{X}^\star(\vec{g})}        &\qw &\nogate{\qbra{\vec{m}_\sat''}} \qw & 
    }
    }
    \end{split}.
    \end{align}

The above describes the CZ network at a single size-$n$ macronode. Not shown in the circuit is that stitching and reduction has also been performed at all the other macronodes. Grouping the star-type CZ networks at each node results in the CZ network $\CZ(\mat{A}_G)$ entangling only the central modes, with a star-type CX network at each macronode entangling the modes there. Critically, for integer-weight $\vec{g}$ and $\mat{A}$, these networks act as logical operators on square-lattice GKP states, and the CX networks vanish on the input $\ket{+}$ states, completely disentangling the satellite modes from the central modes. 

The derivations in this Supplemental Material proceed assuming the central modes are measured; however, stitching and reduction do not rely on this. Consider the case of integer $\vec{g}$ and $\mat{A}$, where $N$ central modes in some macronization of $G$ are left unmeasured. After stitching and reduction, one obtains
  \begin{align}
  \label{circ:disp_squeeze}
    \Qcircuit @C=0.3cm @R=0.5cm {
     \nogate{\ket{+}^{\otimes N}} &{/} \qw &\gate{C_\text{Z}(\mat{A}_G)}  &\gate{S(\vec{\zeta})} & \gate{D(\mathbf{z})} & \qw
    }
    \end{align}
This is a square-lattice GKP graph state $\ket{G}$ up to local unitaries: a terminal squeezing at each central mode and correlated displacements, described by $\mathbf{z}$, due to measured satellite modes at each macronode.

\subsection{Projective Gaussian measurements as rescaled homodyne measurements}
\label{appsec:Gaussmeas}

Often we wish to measure the Gaussian-transformed quadrature $\hat U'^\dag \op q  \hat U' = \alpha \op{q} + \beta \op{p}$, described by projections onto left eigenstates, $\qbra{k} \op{U}$. We show here how to perform this measurement through homodyne detection and a rescaling of the measurement outcome.

First, use a pre-Iwasawa decomposition~\cite{ma1990, Arvind1995, Houde2024} to decompose the single-mode Gaussian unitary $U$ into a shear, squeezing, and a phase delay: 
    \begin{equation} \label{eq:preIwasaza}
        \op{U} = \op{P}(\sigma) \op{S}(\zeta) \op{R}(\beta),
    \end{equation}
where $\op{P}(\sigma) \coloneqq e^{i \frac{\sigma}{2} \op{q}^2}$.
Applying this decomposition to the left eigenstates, the shear acts trivially, giving
\begin{align} \label{eq:simmeasurement}
    \brasub{k}{q} \op{U} 
    = \brasub{\zeta^{-1} k}{q} \op{R}(\beta)
    = \brasub{k'}{q_\beta}.
\end{align}
Alternatively, the above relation above can be written
    \begin{align}
    \label{eq:iwasawa_shifts}
       \brasub{0}{q} \op{U} \op{R}^\dagger(\beta) \op{D}(\tfrac{1}{\sqrt{2}} \mat{S}_R \mat{S}_U^{-1} \tilde{\vec{k}} )  
       = 
       \qbra{k'} 
    \end{align}
where $\tilde{\vec{k}} = [k \,\, 0]^\tp$. Using $\brasub{0}{q} \op{U} = \brasub{0}{q} \op{R}(\beta)$, this reveals that $k'$ can also be expressed as the first entry of $\mat{S}_R \mat{S}_U^{-1} \tilde{\vec{k}}$. In other words, a measurement of $\hat U^\dag \op q  \hat U$ with outcome $k$ is achieved through homodyne detection of a rotated quadrature $\op{q}_\beta \coloneqq \cos \beta \op{q} - \sin \beta \op{p}$ with a rescaling of the outcome.

\subsection{GKP qubit measurements on the central modes}
\label{appsec:logmeas}

Based on the derivation in Sec.~\ref{appsec:Gaussmeas}, here we describe the physical measurement and outcome rescaling one needs to perform to effect the desired homodyne measurement on the canonical graph state. Equating the physical (raw homodyne) circuits at the left-hand-side of~\eqref{appeq:stitcherreduction} and~\eqref{circ:disp_squeeze}, and the canonical (qubit GKP) circuit~\eqref{appeq:stitcherreduction}, we have that
\begin{align}
    \brasub{m_C}{q_\theta} \op{D}(z) \op{S}(\zeta) =  \brasub{m''_C}{q_{\theta'}}
\end{align}
where $\op{q}_{\theta'}$ is the desired rotation. For example, $\theta' \in \{ \frac{\pi}{2} , \pm \frac{\pi}{4}, 0 \}$ realize square-lattice GKP $X, Y,$ and $Z$ measurements, respectively. This means that
\begin{align}
    \brasub{m_C}{q_\theta}\op{D}(z)  = \brasub{m_C''}{q} \op{R}(\theta') \op{S}^\dagger(\zeta).
\end{align}
We can now perform the pre-Iwasawa decomposition, Eq.~\eqref{eq:preIwasaza}, on $\op{U} = \op{R}(\theta') \op{S}^\dagger(\zeta)$, and conclude that the physical $\theta$ ought to be the rotation in the last term of the decomposition. Using expressions in Ref.~\cite{Houde2024}, the correct choice is
\begin{align}
    \theta = \cos^{-1}{\left( \frac{\cos{\theta'}}{ \sqrt{ \cos^2{\theta'} +  \zeta^4\sin^2{\theta'}}} \right)},
\end{align}
which reduces to $\theta'$ for no squeezing ($\zeta = 1$). Square-lattice logical GKP measurements are therefore implemented by measurement angles $\theta' = (\tfrac{\pi}{2}, \cos^{-1}(\sqrt{1+\zeta^4}^{-1}), 0)$, revealing that only $Y$ measurements differ.

\subsection{Shift matrices} \label{appsec:shiftmatrices}

The reduction in the above sections corresponds to the following equivalence
\begin{align} \label{eq:shiftcomparison}
    \qbra{\vec{m}} \op{R}(\vec{\theta}_C) \op{U}_\text{split} \CZ(\mat{A}_\text{DB}) = \qbra{\vec{m}''} \op{R}(\vec{\theta}_L) \CZ(\mat{A}_G) \CX(\vec{g}_T) 
\end{align}
where $\vec{m}$ are the raw homodyne outcomes obtained from the passive circuit, $\vec{m}''$ are the processed outcomes to be interpreted according to a canonical lattice, and $\mat A_\text{DB}$ is the adjacency matrix corresponding to a set of dumbbells. $\CX(\vec{g}_T)$ is the global CX network comprised of the $\CX^\star(\vec{g})$ local to each macronode. $\vec{\theta}_C$ is a list of measurement angles for the central modes, and $\vec{\theta}_L$ are the desired logical angles on the reduced circuit---see Sec.~\ref{appsec:logmeas}.
When and $\vec{g}$ and $\vec{h}$ contain only integers, the canonical lattice corresponds to a square-lattice GKP graph state. 

By extracting the raw measurement outcomes into a shift vector $\tilde{\vec{m}} = [\vec{m} \,\, \vec{0}]^\tp \in \mathbb{R}^{2n}$, we can transform them using Eq.~\eqref{eq:displacementsymplecticrule} to find the simulated outcomes,
\begin{align} \label{eq:shiftmatrices}
    \tilde{\vec m}'' = \mat{S}_{L} \mat S_{G} \mat S_{\text{CX}} \mat S_\text{DB}^{-1} \mat S_\text{split}^{-1} \mat S_{R_C}^{-1} \tilde{\vec m}.
\end{align}
The first $n$ entries of $\tilde{\vec{m}}''$ describe the canonical outcomes $\vec{m}''$ on the right side of Eq.~\eqref{eq:shiftcomparison}.
Note that the final transformation in Eq.~\eqref{eq:shiftmatrices} is not critical, but it is convenient---$\mat{S}_{L}$, corresponding to $\op{R}(\theta_L)$, moves the logical information to the entries in $\tilde{\vec{m}}''$ corresponding to the central modes.

\section{Splitter designs}
\label{ap:alt_splitters}

Infinitely many different splitters can be reduced to the same integer-weight star-type CX network using position measurements following the prescription in the main text. Here we detail several useful splitters using circuit-level descriptions, showing that each produces a weight-1 star-type network of CX gates when its beam splitters are properly tuned. For a given macronode size, each splitter produces the same state and enjoys the same noise properties (see Sec.~\ref{ap:noise_propagation}). At the logical level, the defining difference between them is the shift rules for combining the measurement outcomes. At the physical level, each involves a a different arrangement of beam splitters with different transmissivities. In a practical setting, the best version may be influenced by experimental considerations, such as accumulated noise (chiefly loss) due to varying the optical depths, i.e. the propagation length and beam splitter depth as a function of the input size. Furthermore, despite the splitter's behaviour being equivalent in the homogeneous noise models that we consider, in practice each one may have varying sensitivity to imperfections such as non-uniform losses, fluctuations in beam splitter angles, and phase noise. For the present discussion we focus primarily on the difference in depth of the optical circuit.

The circuits in this section differs slightly from those previously. The measurement outcomes on the satellite modes are set to zero, allowing us to ignore classical shifts. Furthermore, the central modes are left unmeasured to emphasize that, while the shift matrix relies on the particular measurement choice, the reduction to logical graph is independent of the central mode's measurement for these splitters. Finally, we preserve the \emph{terminal squeezing} $S(\zeta)$, which determines the proper quadrature to measure to simulate GKP Pauli measurements and is the factor by which the central mode's measurement outcome is ultimately rescaled. 
In Sec.~\ref{ap:noise_propagation}, we see that the terminal squeezing is a proxy for the severity of noise amplification due to the stitcher. We find that that this factor---$\sqrt{n}$ for valence $n$---is the same for every splitter.

\subsection{Star splitter}\label{sec:direct_splitter}

A star splitter over $n$ modes has $n-1$ beam splitters coupling the central mode (mode 1) to each satellite mode,
    \begin{align}
        \hat{U}^{(n)}_\text{star} = \prod_{k=1}^{n-1} \hat{B}_{1,k+1}(\theta_{k}),
    \end{align}
mimicking the structure in the desired CX network. In the expression above, beam splitters with lower $k$ act first; this is necessary to specify due to the non-commutativity of beam splitters.
For this splitter, the total number of beam splitters, $n-1$, is linear in the number of modes. However, the the central mode sees all $n$ beam splitters, while each satellite mode sees only one. A four-mode example is
\begin{align}\label{bsNetworkstarsplit}
    \begin{split}
        \raisebox{-3em}{
        \hspace{-5em}
        $\hat{U}_\text{star}^{(4)} =$ \quad
        }
        \Qcircuit @C=0.7cm @R=0.7cm
        {
          &\varbs{1}[\theta_{1}] &\varbs{2}[\theta_{2}][0.75] &\varbs{3}[\theta_{3}][0.82] &\rstick{\text{(central mode)}}\qw \\
          &\qw                   &\qw                   &\qw                   &\qw \\
          &\qw                   &\qw                   &\qw                   &\qw \\
          &\qw                   &\qw                   &\qw                   &\qw                    
        }
    \end{split} 
\end{align}

To reduce the circuit to a network of CX gates, we use Eq.~\eqref{eq:BSdecomp_qmeas} on each measured mode and push the squeezing on the central mode through all the CX gates using Eq.~\eqref{eq:squeezingcontrolledgate}. For $n$ modes, the result is the circuit
\begin{align}\label{CXNetworkdirect}
    \begin{split}
        \Qcircuit @C=0.4cm @R=0.4cm
        {
          &\ctrlg{g_1}{1} &\ctrlg{g_2}{2}[0.76] &\qw &\qwdot &\ctrlg{g_{n-1}}{4}[0.8] &\gate{S(\zeta)}      &\qw \\
          &\targ          &\qw            &\qw &\qwdot &\qw            &\rstick{\qbra{0}} \qw& \\
          &\qw            &\targ          &\qw &\qwdot &\qw            &\rstick{\qbra{0}} \qw& \\
          &               &\vdots         &   &\qwdot &               &                     & \\
          &\qw            &\qw            &\qw &\qwdot &\targ          &\rstick{\qbra{0}} \qw&
        }
    \end{split} 
\end{align}
with weights  $g_k = \tan \theta_k(\prod_{j=1}^{k-1} \sec \theta_j)$, and terminal squeezing $\zeta = \prod_{k=1}^{n-1} \sec \theta_k$.

The beam splitter transmissivities for $g_k = -1$ can be obtained through a recursion relation which results in
    \begin{align}
        \theta_{k} = -\tan^{-1} \sqrt{1/k}. 
        \, 
    \end{align}
Using $\sec \,(-\tan^{-1} x) = \sqrt{1+x^2}$, we find $\sec \theta_k = \sqrt{1 + \frac{1}{k}}$, which gives terminal squeezing of $\sqrt{n}$.

\subsection{Cascade splitter}\label{sec:linear_splitter}

A cascade splitter over $n$ modes has $n-1$ beam splitters, one from each mode to the next, 
    \begin{align}
        \hat{U}^{(n)}_\text{cascade} = \prod_{k=1}^{n-1} \hat{B}_{n-k+1,n-k}(\theta_{k}),
    \end{align}
again with beam splitters of lower $k$ acting first. This gives a depth $n-1$ beam splitter network, just like the direct linear splitter. In this case, each modes sees 1 (the central and final mode) or 2 (the rest) beam splitters. The top mode ($k=1$) serves as the central mode. A four-mode example is
\begin{align}\label{ap:bsNetworkcascade}
    \begin{split}
        \raisebox{-3em}{
        \hspace{-5em}
        $\hat{U}_\text{cascade}^{(4)} =$ \quad
        }
        \Qcircuit @C=0.7cm @R=0.7cm
        {
          &\qw                    &\qw                         &\varbs{1}[\theta_{3}] &\rstick{\text{(central mode)}}\qw \\
          &\qw                    &\varbs{1}[\theta_{2}]                         &\qw                   &\qw      \\
          &\varbs{1}[\theta_{1}]                    &\qw      &\qw                   &\qw \\
          &\qw &\qw                         &\qw                   &\qw                    
        }
    \end{split} \; .
\end{align}

To reduce the circuit, use Eq.~\eqref{eq:BSdecomp_qmeas} on each measured mode. Then push the squeezing on each mode through the CX gate and onto the measurement using Eq.~\eqref{eq:squeezingcontrolledgate}. Only the terminal squeezing on the central mode survives. For $n$ modes, the result is the circuit
\begin{align}\label{CXNetworkcascade}
    \begin{split}
        \Qcircuit @C=0.4cm @R=0.4cm
        {
          &\qw                &\qw   &\qwdot &\qw            &\ctrlg{g_{n-1}}{1} &\gate{S(\zeta)}       &\qw \\
          &\qw                &\qw   &\qwdot &\ctrlg{g_{n-2}}{1} &\targ          &\rstick{\qbra{0}} \qw & \\
          &\qw                &\qw   &\qwdot &\targ          &\qw            &\rstick{\qbra{0}} \qw & \\
          &                   &      &\qwdot &\vdots         &               &                      & \\
          &\ctrlg{g_{1}}{1} &\targ &\qwdot &\qw            &\qw            &\rstick{\qbra{0}} \qw & \\
          &\targ              &\qw   &\qwdot &\qw            &\qw            &\rstick{\qbra{0}} \qw &
        }
    \end{split} 
\end{align}
with weights $g_{1} = \tan \theta_{1}$ and $g_{k} = \tan \theta_k / \sec \theta_{k-1}$ for $1<k\leq n-1$,
and the terminal squeezing is $\zeta = \sec \theta_{n-1}$.

Setting $g_k = -1$ and solving the recursion relation gives
    \begin{equation}
        \theta_{k} = -\tan^{-1} \sqrt{k}
        \,  ,
    \end{equation}
with $\sec \theta_k = \sqrt{k+1}$. Just like the other splitters, the terminal squeezing is $\zeta = \sqrt{n}$.

One more step remains. The reduced CX network above is not a star-type network; however, it is equivalent to one. To see this, commute any CX gates whose control is on a measured mode towards the measurements using Eq.~\eqref{commuteCX}. All gates whose control is measured will vanish, leaving only gates in a star-type CX network, albeit with new weights on each gate. 

When all $g_{k} = - 1 $, the weights on any new CX gates created in this process will also be weight $- 1$ by virtue of Eq.~\eqref{commuteCX}, and the desired star-type CX network with $\vec{g} = -\vec{1}$ is generated. 
As CX gates of weight $\pm 1$ have identical effect on square-lattice GKP states; 
the only difference is in the shift matrices, Eq.~\eqref{eq:shiftmatrices}.

\subsection{Tree splitter}\label{sec:tree_splitter}

A \emph{tree splitter}, for $n$ modes, $\op U_\text{tree}^{(n)}$, contains $n-1$ beam splitters. Although the total number of beam splitters is linear in the number of modes, the circuit is log depth, $\log_2 n$, making the tree splitter more compact than the others considered here. The beam splitter network is
\begin{align}\label{eq:tree_splitter}
    \op U_\text{tree}^{(n)} \coloneqq & \prod_{\ell=1}^{\lceil\log_2 n\rceil}
    \bigotimes_{k=0}^{\lfloor (n-1)/2^\ell \rceil-1} \hat{B}_{2^\ell k+1,2^\ell (k+1/2)+1}(\theta_{\ell,k}),
\end{align}
with the transmissivities labeled by two numbers; $\ell$ specifies the beam splitter layer and $k$ is a counter within that layer.  
The beam splitters layers act on an input state in order: layer $\ell=1$ acts first and layer $\ell=\lceil\log_2 n\rceil$ acts last. Within a layer the beam splitters commute. 
Finally, $\lfloor a \rceil\coloneqq \lfloor a + \tfrac{1}{2}\rfloor$ denotes the nearest integer to a real number $a$.
 
The structure of a tree splitter is well illustrated with a circuit diagram. Here is an example for 7 modes:
\begin{align}\label{bsNetworkCircuitTree}
    \begin{split}
        \raisebox{-5.5em}{
        \hspace{-5em}
        $\hat{U}_\text{tree}^{(7)} =$ \quad
        }
        \Qcircuit @C=0.7cm @R=0.6cm
        {
          &\varbs{1}[\theta_{1,0}] &\qw            &\varbs{2}[\theta_{2,0}][0.25]                       &           \qw                &        \varbs{4}[\theta_{3,0}][0.4]                    &\rstick{\text{(central mode)}} \qw                   
          \\
          &\qw                     &\qw                 &             \qw            &          \qw                 &           \qw                & \qw     
          \\
          &\varbs{1}[\theta_{1,1}] &\qw             &           \qw              &               \qw            & \qw      & \qw                                  
          \\
          &\qw                     &            \qw             &    \qw                     &  \qw  & \qw     & \qw                      
          \\
          &\varbs{1}[\theta_{1,2}] &\qw & \varbs{2}[\theta_{2,1}][0.25] & \qw                       & \qw     & \qw      
          \\
          &\qw & \qw    & \qw                     & \qw                       &        \qw        &\qw     
          \\
          &\qw                     & \qw                     & \qw                     &  \qw & \qw &\qw
        }
    \end{split}
\end{align}
This example illustrates another feature: any tree splitter can be decomposed into two smaller tree splitters (allowing for tree splitters of size 1) connected by a single beam splitter,
    \begin{align}\label{eq:tree_splitter_decomp}
    \op U_\text{tree}^{(n)} 
    =
    \op U_\text{tree}^{(k)} \otimes \op U_\text{tree}^{(k' \leq k)} \op{B}_{1,k+1}(\theta_{\lceil\log_2 n\rceil,0})
\end{align}
with $k + k' = n$. In the above example, $\op{B}_{1,5}$ connects tree splitters of size $k=4$ and $k'=3$, respectively.

To reduce a tree splitter, measure every mode in \emph{q} other than the top mode, which serves as the central mode. Noting that the second mode of each beam splitter is measured without any further connections, we use the identity in Eq.~\eqref{eq:BSdecomp_qmeas} on each measured mode.
Pushing the squeezing operators on the central mode through the CX gates using Eq.~\eqref{eq:squeezingcontrolledgate} gives the circuit
\begin{align}\label{CXNetworkCircuitTree}
    \begin{split}
        \resizebox{0.7\columnwidth}{!}{
        \Qcircuit @C=0.4cm @R=0.4cm
        {
          &\ctrlg{g_{1,0}}{1}         &      \qw            &   \ctrlg{g_{2,0}}{2}[0.28]                    &           \qw                &        \ctrlg{g_{3,0}}{4}[0.15]         &     \qw &  \qwdot         &   \ctrlg{g_{\lceil \log_2 n\rceil,0}}{4}[0.98] &   \gate{S(\zeta)} &\qw&  \qw  
          \\
          &\targ            &        \qw                 &             \qw            &          \qw                 &           \qw            & \qw    & \qwdot  & \qw  &\rstick{\qbra{0}} \qw&&
          \\
          &\ctrlg{g_{1,1}}{1}         &      \qw              &           \targ             &               \qw            & \qw      & \qw          &      \qwdot & \qw  &\rstick{\qbra{0}} \qw&&       
          \\
          &\targ           &            \qw             &    \qw                     &  \qw  & \qw     & \qw        &     \qwdot  & \qw &\rstick{\qbra{0}} \qw&&          
          \\
          &                     &  &  & \vdots                      &      &   & \qwdot  &   &   & 
          \\
          & \ctrlg{g_{1,\lfloor (n-1)/2\rceil - 1}}{1}  & \qw    & \targ                    & \qw    & \qw & \qw &  \qwdot          &\qw  &\rstick{\qbra{0}} \qw&&  
           \\
           &   \targ                     & \qw                    & \qw                     &  \qw & \qw &\qw &\qwdot &\qw &\rstick{\qbra{0}} \qw&&
        }
        }
    \end{split}
\end{align}
with CX weights given by 
\begin{equation}
\begin{split}
    g_{\ell,k}\coloneqq \begin{cases} \tan\theta_{\ell,k} & \ell=1, \\
    \frac{\tan\theta_{\ell,k} \prod_{j=1}^{\ell-1}\sec\theta_{\ell-j,2^jk}}{\prod_{j=1}^{\ell-1}\sec\theta_{\ell-j,2^{j-1}(2k + 1)}}  & \text{Otherwise.}
    \end{cases}
    \end{split}\label{eq:redefinition_theta}
\end{equation}
and terminal squeezing $\zeta = \prod_{\ell=1}^{\lceil \log_2 n\rceil}\sec{\theta}_{\ell,0}$. 

From Eq.~\eqref{eq:redefinition_theta}, setting  $g_{\ell,k} = -1$ recursively determines the original beam splitter angle $\theta_{\ell,k}$:
\begin{align} \label{eq:treesplitterangles}
    \theta_{\ell,k} = -\tan^{-1} \Big( \sqrt{t_{\ell,k}/c_{\ell,k} } \Big)
\end{align}
with $\sec \theta_{\ell,k} = \sqrt{1 + t_{\ell,k}/c_{\ell,k}}$. 
Here, $c_{\ell,k}$ is the number of modes in the subtree connected to the control mode of the CX gate, and $t_{\ell,k}$ is the number of modes connected to the subtree of the target mode. One may refer directly to the original beam-splitter circuit to find these numbers.
Note that the first layer, $\ell=1$, are always 50:50 beam splitters, $\theta = \frac{\pi}{4}$ and that the terminal squeezing on the central mode is $\zeta= \sqrt{n}.$
The final step in the reduction is to recognize that the CX network in Eq.~\eqref{CXNetworkCircuitTree} is equivalent to a weight $\vec{g} = -\vec{1}$ star-type CX network when $ g_k = -1 $. This equivalence is described above for the cascade splitter. 

For clarity, consider the 7-mode tree splitter in Eq.~\eqref{bsNetworkCircuitTree}. The terminal squeezing is $\zeta = \sqrt{7}$. The top mode of beam splitter $\op{B}_{3,0}$ is connected to a subtree with four modes, $c_{3,0} = 4$, and the bottom mode to a subtree with three modes, $t_{3,0} = 3$, so its transmissivity is $\theta_{0,3} = -\tan^{-1} \sqrt{3/4}$. Similarly, $\theta_{1,0} = \theta_{1,1} = \theta_{1,2} = \theta_{2,0} = -\tan^{-1} 1 = -\frac{\pi}{4}$ and $\theta_{2,1} = -\tan^{-1} \frac{1}{\sqrt{2}}$.

\subsection{\texorpdfstring{$2^j$}{2**j} splitter}\label{sec:2N_splitter}

In the above splitters, the central mode is pre-designated (as the top mode). Here, we present the $2^j$ splitter, $\op U_{2^j}^{(n)}$, with linear depth $n-1$, for which \emph{any} of the modes can be chosen as the central mode, and the others are measured in $\op{q}$. The cost for this freedom is that this splitter can only stitch macronodes with $n = 2^j$ modes for integer $j$, and the number of beam splitters is large: $j 2^{j-1} = n \log_2 \sqrt{n}$.

The $2^j$ splitter network is constructed by first coupling a pair of modes with a balanced beam splitter. Then, each mode of that pair is coupled to its partner mode in another pair with another balanced beam splitters. Repeating this procedure doubles the number of modes at each step, creating a network similar to the tree splitter but with more modes and more beam splitters (all of them balanced):
\begin{align}\label{eq:2Nsplitter}
    \op U^{(n)}_{2^j} 
    =
    \prod_{\ell=1}^{\log_2 n}
    \bigotimes_{k'=1}^{2^{\ell-1}}
    \bigotimes_{k=0}^{n/2^\ell - 1}
    \hat{B}_{2^\ell k+k',2^\ell (k+1/2)+k'} \big(\tfrac{\pi}{4} \big),
\end{align}
An example for $n=8$ modes ($j=3$) is
\begin{align}
\begin{split}
\raisebox{-6em}{$
\op U^{(8)}_{2^j} = $
}
\quad
\Qcircuit @C=0.6cm @R=0.6cm
{
& \bsbal{1} & \bsbal{2} & \qw & \bsbal{4} & \qw & \qw & \qw & \qw\\
& \qw & \qw & \bsbal{2} & \qw & \bsbal{4} & \qw & \qw & \qw\\
& \bsbal{1} & \qw & \qw & \qw & \qw & \bsbal{4} & \qw & \qw\\
&\qw & \qw & \qw & \qw & \qw & \qw & \bsbal{4} & \qw\\
& \bsbal{1} & \bsbal{2} & \qw & \qw & \qw & \qw & \qw & \qw\\
& \qw & \qw & \bsbal{2} & \qw & \qw & \qw & \qw & \qw \\
& \bsbal{1} & \qw & \qw & \qw & \qw & \qw & \qw & \qw \\
&\qw & \qw & \qw & \qw & \qw & \qw & \qw & \qw \\
}
\end{split}\,.
\end{align}

To reduce a $2^j$ splitter, first choose a mode to serve as the central mode. Then, measure all other satellites mode in \emph{q}. No matter the choice for the central mode, many of the beam splitters vanish due to the identity~\cite{Walshe2023equivalent},
\begin{align}\label{eq:BSvanish}
    \begin{split}
        \Qcircuit @C=0.5cm @R=1cm
        {
         &\varbs{1} &\rstick{\qbra{0}} \qw                   \\
         &\qw       &\rstick{\qbra{0}} \qw 
        }
        \quad \quad \quad \raisebox{-1.5em}{=} \quad
       \Qcircuit @C=0.5cm @R=1cm
        {
         &\qw &\rstick{\qbra{0}} \qw                   \\
         &\qw &\rstick{\qbra{0}} \qw 
        }
    \end{split}
\end{align}
The remaining beam splitters form a tree splitter up to a permutation, with the unmeasured mode of the $2^j$ splitter corresponding to the central mode of the tree splitter.
From this perspective, the control mode and target mode corresponding to the final beam splitter connect subtrees with the same number of modes in any layer. We can use the same reduction as that for the tree splitter with $\theta=-\tan^{-1}(1) = -\tfrac{\pi}{4}$ for all the beam splitters.  The terminal squeezing $\zeta$ is thus $\sqrt{n}$.

As a final note: the four splitter that formed the backbone of the stitching for the RHG lattice in Ref.~\cite{tzitrin2021fault} is a $2^j$ splitter with $j=2$, and it is equivalent to an $n=4$ tree splitter.
\section{Accommodating arbitrary GKP lattices and anisotropic noise}
\label{ap:gkp_lattice}

Here we elaborate on our results for incorporating arbitrary-lattice GKP states with anisotropic noise into our passive architectures. We will be propagating the GRN channel $\mathcal{E}$ through stitching circuits, relying on the $\mat X/ \mat Y$ matrix picture of the action of completely positive Gaussian maps described in Sec.~\ref{ap:notation}.

Assume the availability of GKP states $\op{\latG}\ket{\varnothing}$ followed by GRN channel $\mathcal{E}$. Because any two identical single-mode Gaussian unitary operations commute past a beam splitter, we obtain the following equivalent circuits:
\begin{align}
    \label{appeq:noisy-dumbells}
    \begin{split}
    \resizebox*{0.8\columnwidth}{!}{
    \Qcircuit @C=0.3cm @R=0.4cm { 
    \nogate{ \mathcal E \op{\latG} \ket{\qunaught}} & \bsbal{1} & \qw & \qw & \nomultigate{1}{=}&  \nogate{\ket{+}} & \ctrl{1} & \gate{\mathcal E \latG} & \nogate{A} \qw  
    \\
     \nogate{  \mathcal E \op{\latG} \ket{\qunaught}} & \qw & \gate{F} & \qw & \nghost{=}& \nogate{\ket{+}} & \ctrl{-1} & \gate{F \mathcal E \latG  F^\dag} & \nogate{B} \qw 
    } 
    }
    \end{split}  ,
\end{align}
labeling the two halves as A and B to distinguish their different noise properties.
    \begin{equation}
    \begin{split}
    \includegraphics[width=.25\textwidth]{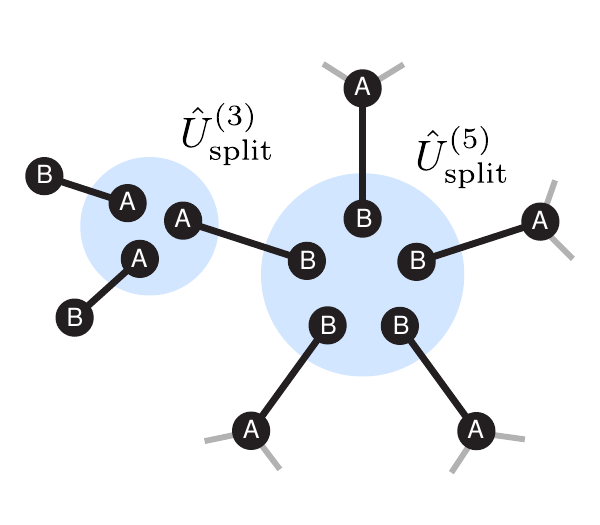}
    \end{split}.
    \end{equation}
The sub-circuit for the splitter coupling type-A halves is
\begin{align}
    \begin{split}\label{eq:totl1}
    \resizebox*{0.75\columnwidth}{!}{
    \Qcircuit @C=0.3cm @R=0.5cm {
     &\nogate{\ket{+}} &\ctrl{3} &\qw      &\qw      &\qw       &\multigate{2}{U_\text{split}^{(3)}} &\gate{\mathcal E \latG} &  \nogate{_{\theta_{\cent, A}}\!\!\bra{m_{\cent, A}}} \qw                        
     \\
     &\nogate{\ket{+}} &\qw      &\ctrl{3} &\qw      &\qw       &\ghost{U_\text{split}^{(n)}}     &\gate{\mathcal E \latG} & \nogate{_{\theta_{\sat, A}}\!\!\bra{m_{\sat, A,2}}} \qw                                                   
     \\
     &\nogate{\ket{+}} &\qw      &\qw      &\ctrl{3}     &\qw       &\ghost{U_\text{split}^{(n)}}     &\gate{\mathcal E \latG} & \nogate{_{\theta_{\sat, A}}\!\!\bra{m_{\sat, A, 3}}} \qw                                                   
     \\
     &\nogate{\ket{+}} &\ctrl{0} &\qw      &\qw      &\qw       &\gate{F \mathcal E \latG F^\dag}          &   \qw                
     \\
     &\nogate{\ket{+}} &\qw      &\ctrl{0}      &\qw      &\qw       &\gate{F \mathcal E \latG F^\dag}          &   \qw                
     \\
     &\nogate{\ket{+}} &\qw      &\qw &\ctrl{0}      &\qw       &\gate{F \mathcal E \latG F^\dag}          &  \qw               
     \relax \inputgroupv{1}{3}{.1em}{1.5em}{\raisebox{-1.35cm}{A}}
     \relax \inputgroupv{4}{6}{.1em}{1.5em}{\raisebox{-1.35cm}{B}}
    }
    }
    \end{split},
\end{align}
where we have a slight change in notation: $\brasub{m_{\cent, A}}{\theta_{C, A}}$ and $\brasub{m_{\sat, A}}{\theta_{S, A}}$ denote central- and satellite-mode measurements for macronodes of type-A with basis $\op{q}_{\theta_{\ldots}}$.

Now, we would like to undo the effect of the lattice transformation and the Fourier transforms, while ensuring that the pre-measurement noise channel remains GRN. This is achieved by choosing $\theta$'s corresponding to the projection $\brasub{\cdot}{\beta_{\_, A}} \op{V}^\dagger$ for modes of type-A, and $\brasub{\cdot}{\beta_{\_, B}}\op{F} \op{V}^\dagger \op{F}^\dagger$ for modes of type-B, where the angles $\beta$ correspond to the measurement settings when $\op{V} = \op{I}$. With this choice, the effective noise in the circuit is modified:
\begin{equation}
\label{appeq:newnoise}
     \resizebox{0.85\columnwidth}{!}{$
     \begin{aligned}
        \Qcircuit @C=0.3cm @R=0.5cm 
    { 
    &\qw & \gate{\mathcal E \latG} & \nogate{\brasub{m_{\_, A}}{\theta_{\_, A}}} \qw}
    &=
    \Qcircuit @C=0.3cm @R=0.5cm 
    {
    &\qw &\gate{\latG^\dag \mathcal E \latG} & \nogate{\brasub{\cdot}{\beta_{\_, A}}} \qw 
     } 
     \\
    \Qcircuit @C=0.3cm @R=0.5cm 
    { 
    &\qw & \gate{F \mathcal E \latG F^\dag} & \nogate{\brasub{m_{\_, B}}{\theta_{\_, B}}} \qw}
    &=
    \Qcircuit @C=0.3cm @R=0.5cm 
    {
    &\qw &\gate{F \latG^\dag \mathcal E \latG F^\dag} & \nogate{\brasub{\cdot}{\beta_{\_, B}}} \qw 
     }
     \end{aligned}
     $},
\end{equation}
Furthermore, as in Sec.~\ref{appsec:Gaussmeas} and~\ref{appsec:logmeas}, we would like to determine the physical $\theta$'s for which the circuit performs the desired GKP qubit measurement on the canonical graph. We can do this by choosing the rotation angle in the pre-Iwasawa decomposition of the following operators:
\begin{align}
    \text{A satellites}: & \quad \op{V^\dagger} \\
    \text{B satellites}: & \quad \op{F} \op{V^\dagger} \op{F}^\dagger \\
    \text{A central}: & \quad \op{R}(\theta_L) \op{S}^\dagger(\sqrt{n}) \op{V^\dagger}\\
    \text{B central}: &  \quad \op{R}(\theta_L) \op{S}^\dagger(\sqrt{n}) \op{F} \op{V^\dagger} \op{F}^\dagger,
\end{align}
with the rotation and squeezing terms the same as in~~\ref{appsec:logmeas}.

We can confirm that the pre-measurement noise channel is GRN. The input GRN channel $\mathcal{E}$ is described by matrices $\mat{X} = \mat{I}$ and $\mat{Y} = \mat{\Sigma}_\mathcal{E}$---see Eq.~\eqref{appeq:Gaussianchannel}. The updated noise on type-A and type-B macronodes is obtained by composing the input channel with the $\mat{X}$ and $\mat{Y}$ matrices of $\op{V}$ and $\op{F}$:
\begin{align}
 \label{eq:epsilon_a}
    \mathcal{E}_A &\coloneqq \op{\latG}^\dagger\mathcal{E}\op{\latG} {} : 
    \begin{cases} \mat X_A 
    =  \mat S^{-1}_\latG \mat I \mat S_\latG  
    = \mat I \\ 
    \mat  Y_A =  \mat S^{-1}_\latG \mat \Sigma_\mathcal{E} \mat S^{-\tp}_\latG  \end{cases} 
    \\
    \label{eq:epsilon_b}
    \mathcal{E}_B &\coloneqq \hat{F} \op{\latG}^\dagger\mathcal{E}\op{\latG} \hat{F}^\dagger{} : 
    \begin{cases} \mat X_B 
    = \mat S_F \mat S^{-1}_\latG \mat I \mat S_\latG \mat S^{-1}_F 
    = \mat I \\ 
    \mat  Y_B = \mat S_F \mat S^{-1}_\latG \mat \Sigma_\mathcal{E} \mat S^{-\tp}_\latG \mat S^\tp_F \end{cases}. 
\end{align}
Since $\mat X_A = \mat{X}_B = \mat{I}$, the modified noises are also additive GRN channels, but with different covariance matrices. 

Finally, a shift matrix modified relative to~\eqref{eq:shiftcomparison} accommodates the transformations above to find the processed outcomes $\vec{m}''$ from the measured outcomes $\vec{m}$,
\begin{align} \label{eq:TOTLshiftmatrices}
    \tilde{\vec m}'' = \mat{S}_{L} \mat S_{G} \mat S_{\text{CX}} \mat S_{\text{DB}}^{-1} \mat S_\text{split}^{-1} \mat{S}^{-1}_B \mat{S}^{-1}_A \mat S_{R_C}^{-1} \tilde{\vec m},
\end{align}
where $\mat{S}^{-1}_A$ and $\mat{S}^{-1}_B$ are the symplectic matrices associated with the local transformations $\op{\latG}$ and $\op{F} \op{\latG} \op{F}^\dagger$ on the type-A and type-B modes, respectively (each matrix only has support on those modes of the associated type).

To summarize, we have shown that, given non-square-lattice GKP dumbbells, one can find an appropriate homodyne measurement setting to simulate the desired measurement on a cluster state constructed with square-lattice GKP states (with modified effective noise). Notice that, when $\op{\latG}$ is the identity and the noise is isotropic, we recover the standard square-lattice version of the architecture.

\section{Leveraging quadrature bias and proof of noise decoupling}
\label{ap:proof_noise_decouple}

\begin{claim}
Suppose that we have an architecture with:
\begin{itemize}
    \item Uniformly rectangular GKP states with uniform, uncorrelated Gaussian Random Noise;
    \item Entanglement generated through dumbbell-splitters as above;
    \item A target graph state with bipartition into modes $A$ and $B$;
\end{itemize}

Then, the phase error rates associated with $X$ measurements on nodes $A$ depends only on the noise along one quadrature of the input states, and those along the nodes $B$ depend on noise along the other quadrature.
\end{claim}

\begin{proof}
Let us consider the family of rectangular GKP states which have the associated lattice transformation $\mat{S}_\alpha = \text{diag}[\alpha, \frac{1}{\alpha}]$ for some $\alpha \in \mathbb{R} $. As above, non-ideal rectangular GKP states can be modelled by ideal GKP qunaught states followed first by $\mat S_\alpha$ and then by GRN. If the noise between $\hat q$ and $\hat p$ quadratures is uncorrelated, we can interpret the noise covariances $\epsilon_{q(p)}$ as the $\hat q(\hat p)$ variances of each of the (infinitely many) GKP peaks in our model. We can furthermore bridge our model with other figures of merit for GKP states by setting $\epsilon_{q(p)}$ equal, for example, to the effective squeezing~\cite{Duivenvoorden2017} $\sigma^2_{\alpha, q(p)}$ with respect to the lattice specified by $\alpha$. From Eqs.~\eqref{eq:epsilon_a} and~\eqref{eq:epsilon_b}, we know that this covariance matrix gets updated to $\mat \Sigma_{A(B)}$, associated with GRN channel; $\mathcal{E}_{A(B)}$ on macronodes A(B), where $\mat \Sigma_A = \text{diag}[\epsilon_q/\alpha^2, \alpha^2\epsilon_p]$ and $\mat \Sigma_B = \text{diag}[\alpha^2\epsilon_p, \epsilon_q/\alpha^2]$. This form of the covariance matrices shows that both macronodes depend only on the effective squeezing relative to the qunaught state lattice. The Fourier transforms on macronodes $B$ suggest the noise along each quadrature depends on the initial noise along the conjugate quadrature. We have already proven most of the claim; what remains is the phase error probabilities' dependence on the noise along particular quadratures. We have seen from the previous section that $X$ measurements on a particular macronode can be performed by binning the homodyne outcomes of central modes (measured in $\hat p$) and satellite modes of the neighbours (measured in $\hat q$). Furthermore, from Eq.~\eqref{eq:epsilon_a} and ~\eqref{eq:epsilon_b} we know that these measurement bases are unchanged for any rectangular GKP states. Therefore, from $\mat \Sigma_A$ above, we see that a $\hat p$ measurement on the central modes in $A$ depends only on the noise along $\hat p$. Since every neighbour of an $A$ mode is a $B$ mode, a $\hat q$ measurement in $B$ also depends only on noise along $\hat p$ (the first entry in $\mat \Sigma _B$). Thus we have proven the claim for $A$ modes. The argument is symmetrical for $B$ modes, where the errors depend only along the noise in $\hat q$.
\end{proof}

\section{Uniform isotropic noise propagation for square-lattice GKP states}
\label{ap:noise_propagation}

In this section, we examine the effect of noisy states on the effective Pauli measurement results on the canonical cluster state. We show how to calculate the probability of an effective Pauli error both conditioned and unconditioned on the measured homodyne values. For simplicity, we focus on the specific case of square GKP states, $\op{\latG} = \hat{I}$, isotropic Gaussian random noise $\mathcal{E}$ with $\mat{\Sigma}_\mathcal{E} = \epsilon \mat{I}_2$, satisfying $\hat{F}\mathcal{E}\hat{F}^\dagger = \hat{F}^\dagger\mathcal{E}\hat{F} = \mathcal{E}$, and Pauli $X$ measurements on the GKP graph state, which translate to $\op{p}$ measurements on the central modes.

The uniform, isotropic GRN channel commutes through the dumbbell generation circuit, Eq.~\eqref{appeq:noisy-dumbells}. Instead of directly commuting the noise through the splitter at this point, we take a different approach. First, we perform the reduction to $\hat{C}^\star_X(\vec{g})$ (since it is state-independent) and then pull out Fourier transforms to turn it into $\hat{C}^\star_Z(\vec{g})$, which changes the $\op{q}$ measurements into $\op{p}$ measurements, $\qbra{s} = \pbra{s} \op F$. This leaves both the measurement values and the isotropic noise channels unchanged. The circuit for a splitter of size $n$ is
\begin{equation}
\resizebox{0.88\columnwidth}{!}{$
\begin{aligned}
    \Qcircuit @C=0.3cm @R=0.275cm 
    { 
     &\gate{\mathcal E } &\multigate{3}{U_\text{split}^{(n)} } &\nogate{\brasub{m_\cent}{p} }\qw \\
     &\gate{\mathcal E } &\ghost{U_\text{split}^{(n)}}         &\nogate{\qbra{m_2}} \qw\\
     &\nogate{\vdots}    &\nghost{U_\text{split}^{(n)}}         &\nogate{\vdots} \\
     &\gate{\mathcal E } &\ghost{U_\text{split}^{(n)}}         &\nogate{\qbra{m_n}} \qw
    }
     \raisebox{-4em}{=} \;\;
    \Qcircuit @C=0.3cm @R=0.5cm 
    { 
     & \qw & \gate{\mathcal E }  & \ctrlg{g_1}{1} & \qw & \qwdot & \ctrlg{g_n}{3}[0.75]&  \nogate{\brasub{m_C'}{p} }\qw\\
     & \gate{F} & \gate{\mathcal E } & \ctrl{-1} & \qw & \qwdot &\qw &  \nogate{\pbra{m'_2}} \qw\\
     & & \vdots & & & & & \vdots \\
     & \gate{F} & \gate{\mathcal E } & \qw & \qw & \qwdot & \ctrl{-3} &  \nogate{\pbra{m'_n}} \qw
    }
\end{aligned}.
$}
\end{equation}
Pushing the single-mode GRN channels through the $C_Z$ network results in a correlated GRN channel $\mathcal{E}'$ right before the measurements with covariance matrix
$\mat \Sigma_{\mathcal{E}'} = \mat S_{C^\star_Z} \mat \Sigma_{\mathcal{E}^{\otimes n}} \mat S_{C^\star_Z}^\tp$ = $\epsilon \mat S_{C^\star_Z} \mat S_{C^\star_Z}^\tp$, since $\mat \Sigma_{\mathcal{E}^{\otimes n}} = \epsilon \mat I_{2n}$. The symplectic matrix for $\hat{C}^\star_Z(\vec{g})$ is
\begin{equation}
    \mat{S}_{C^\star_Z} = \begin{bmatrix} \mathbf 1 & \mathbf 0 \\ \mat A & \mat 1 \end{bmatrix},
\end{equation}
where $\mat A$ is the weighted adjacency matrix for the graph. By matrix multiplication, we see that the quadrant of $ \mat  \Sigma_{\mathcal{E}'}$ that applies to our $\op{p}$ measurements above, $\vec{m'}$, is $ \mat{K} :=\mat 
 \Sigma_{\mathcal{E}'}^{pp} = \epsilon \left(\mat I_{n} + \mat A^2\right)$. For $\vec{g} = -\vec{1}$, the square of an adjacency matrix simply counts the number of two-step paths leading from one node to another. For a star graph, there are $n-1$ possible two-step paths starting from the central node---each goes to a satellite mode and comes back. There is furthermore just one two-step path joining any pair of distinct and coincident satellite modes. Therefore, 
\begin{align}
\label{eq:covariance}
\mat{K} = \epsilon \begin{bmatrix} 
    n & 0 & 0 & \dots & 0 \\
    0 & 2 & 1 & \dots & 1 \\
    0 & 1 & 2 & \dots & 1 \\
    \vdots &  & & \ddots & \vdots  \\
    0 & 1 & 1 & \dots & 2 \\
\end{bmatrix}
\end{align}
Consequently, independent of the choice of splitter, a $\op{p}$-homodyne measurement of every central mode sees a noise variance of $n \epsilon$, where $n$ is the valence of the corresponding node of the graph. 
That is, varying the splitter $\op{U}^{(n)}_{\text{split}}$ has no impact on fault tolerance for an isotropic GRN noise model.

If measurement values ${\vec{\mathfrak{m}'}}$ were obtained in the absence of noise, with ${\vec{\mathfrak{m}'}} \mod \sqrt{\pi} = 0$, then under the effect of our noise channel, $\vec{m'} = \vec{\mathfrak{m}'} + \vec{x}$, where $\vec{x} \sim \mathcal{N}(\vec{0}, \mat{K})$ is a vector of random variables drawn from a multi-variate normal distribution of means $\vec{0}$ and covariances $\mat{K}$. Standard inner decoding (GKP binning)~\cite{tzitrin2021fault} simply consists in snapping the measurement vector $\vec{m'}$ to $\vec{\bar{\mathfrak{m}}'}$, the nearest vector of integer multiples of $\sqrt{\pi}$ with respect to the Cartesian distance. More intricate binning methods take advantage of the known correlations in $A$ by using the Mahalanobis distance~\cite{Xanadu2024}. The probability of an error in mode $i$ is given by probability of $x_i$ falling outside of $\cup_{j \in \mathbb{Z}} \left[(2j - 1/2)\sqrt{\pi}, (2j + 1/2)\sqrt{\pi} \right]$, which is given by:
\begin{align}
\begin{split}
\text{P}&\left[\frac{\bar{\mathfrak{m}}'_i}{\sqrt{\pi}} \neq \frac{\mathfrak{m}'_i}{\sqrt{\pi}} \! \! \mod 2 \right] \\ & = 1 - \frac{1}{2}\sum_{j = -\infty}^{\infty} \left(\operatorname{erf}\left[(2j + 1/2) \sqrt{\sfrac{\pi}{(2\mat{K}_{ii})}} \right]\right.\\ & \qquad \qquad \qquad- \operatorname{erf}\left.\left[(2j - 1/2) \sqrt{\sfrac{\pi}{(2\mat{K}_{ii})}} \right]\right) \\
& =: f(\mat{K}_{ii})
\end{split}
\end{align}
In the regime of high squeezing, with little overlap between the Gaussians, we have that this expression is well-approximated by:
\begin{equation}
   f(\mat{K}_{ii}) \approx \operatorname{erfc}\left(\frac{\sqrt{\pi}}{2\sqrt{2\mat{K}_{ii}}}\right), \quad \epsilon \to 0,
\end{equation}
where $\operatorname{erfc}$ is the complementary error function.

A canonical Pauli $X$ measurement associated with a given macronode of valence $n$ can be calculated by summing the binned $\vec{m'}$ outcomes from the central node to $k\leq n$ different binned outcomes, each from a satellite mode of a neighbouring macronode. The value $k$ can be lower than $n$ since central nodes do not impart classical shifts to the central nodes of their neighbouring macronodes. This means we can bound the phase or Pauli $Z$ error probability via the union bound as:
\begin{align}
    p^Z_{\text{err}} \leq f(n\epsilon) + k f(2\epsilon)
    \leq  f(n\epsilon)+  n f(2\epsilon) 
\end{align}
The relative impact of the first term tends to grow with valence and decrease with $\epsilon$. For example, with $2 \epsilon = 10.95~\text{dB}$, which is around the threshold for the hyperbolic surface code, the ratio of the first and second term is $2.01$, $3.82$, and $5.43$ for macronodes of valence $3$, $4$, and $5$ respectively.

With the knowledge the measurement values $\vec{m'}$, binned to $\vec{\bar{\mathfrak{m}}'}$, and of the strength of the noise on each of the modes, we can calculate the \textit{conditional} probability that an error may have occurred with the formula
\begin{equation}
\resizebox{0.8\columnwidth}{!}{$
\begin{aligned}
    \text{P}&\left[\frac{\bar{\mathfrak{m}}'_i}{\sqrt{\pi}} \neq \frac{\mathfrak{m}'_i}{\sqrt{\pi}} \!\! \mod 2 \middle| m'_i\right] \\
    &= \frac{\sum\limits_{j=-\infty}^\infty\text{exp}\left[-(m'_i - \bar{\mathfrak{m}}'_i - (2j + 1)\sqrt{\pi})^2 / (2 \mat{K}_{ii})\right]}{\sum\limits_{j=-\infty}^\infty\text{exp}\left[-(m'_i - \bar{\mathfrak{m}}'_i - j \sqrt{\pi})^2/ (2 \mat{K}_{ii})\right]} \\
    &=: g(m'_i - \bar{\mathfrak{m}}'_i, \mat{K}_{ii}).
\end{aligned}
$}
\end{equation}  

To obtain the conditional phase error probability associated with a Pauli $X$ measurement, we need to consider the binned shifts from neighbouring macronodes. Let $\text{neigh}(0)$, with $|\text{neigh}(0)| = k$, be the set of neighbouring macronodes which contribute a binned shift to the canonical $X$ measurement on macronode $0$. Let $m_j^{\prime (i)}$ be the measurement value associated with the mode $j$ of macronode $i$, and let $\text{sat}^{(i)}(0)$, with $i \in \text{neigh}(0)$, be the $j$-index of macronode $i$'s satellite node which is connected to macronode $0$ by a bell pair. Binning errors on an odd number of binned outcomes contributing to the canonical Pauli $X$ measurement would result in a phase error. Using the union-bound, the probability of an error on macronode $0$ can therefore be bounded by:
\begin{align}
\begin{split}
    p^{Z}_{\text{cond}} &\leq g(m_1^{\prime (0)} - \bar{\mathfrak{m}}_1^{\prime (0)}, n\epsilon) \\
    &+ \sum_{i \in \text{neigh}(0)} g(m_{\text{sat}^{(i)}(0)}^{\prime(i)} - \bar{\mathfrak{m}}_{\text{sat}^{(i)}(0)}^{\prime(i)}, 2\epsilon).
\end{split}
\end{align}
This bound on the conditional error probability can be used as soft information in the qubit-level decoder to improve logical error rates and fault-tolerant thresholds. It is straightforward to generalize this analysis to non-anisotropic noise, GKP states on different lattices, and different canonical measurement bases.

\section{Error correction simulation details}
\label{ap:qec_sims}

\begin{figure}
    \centering
    \includegraphics[width=.45\textwidth]{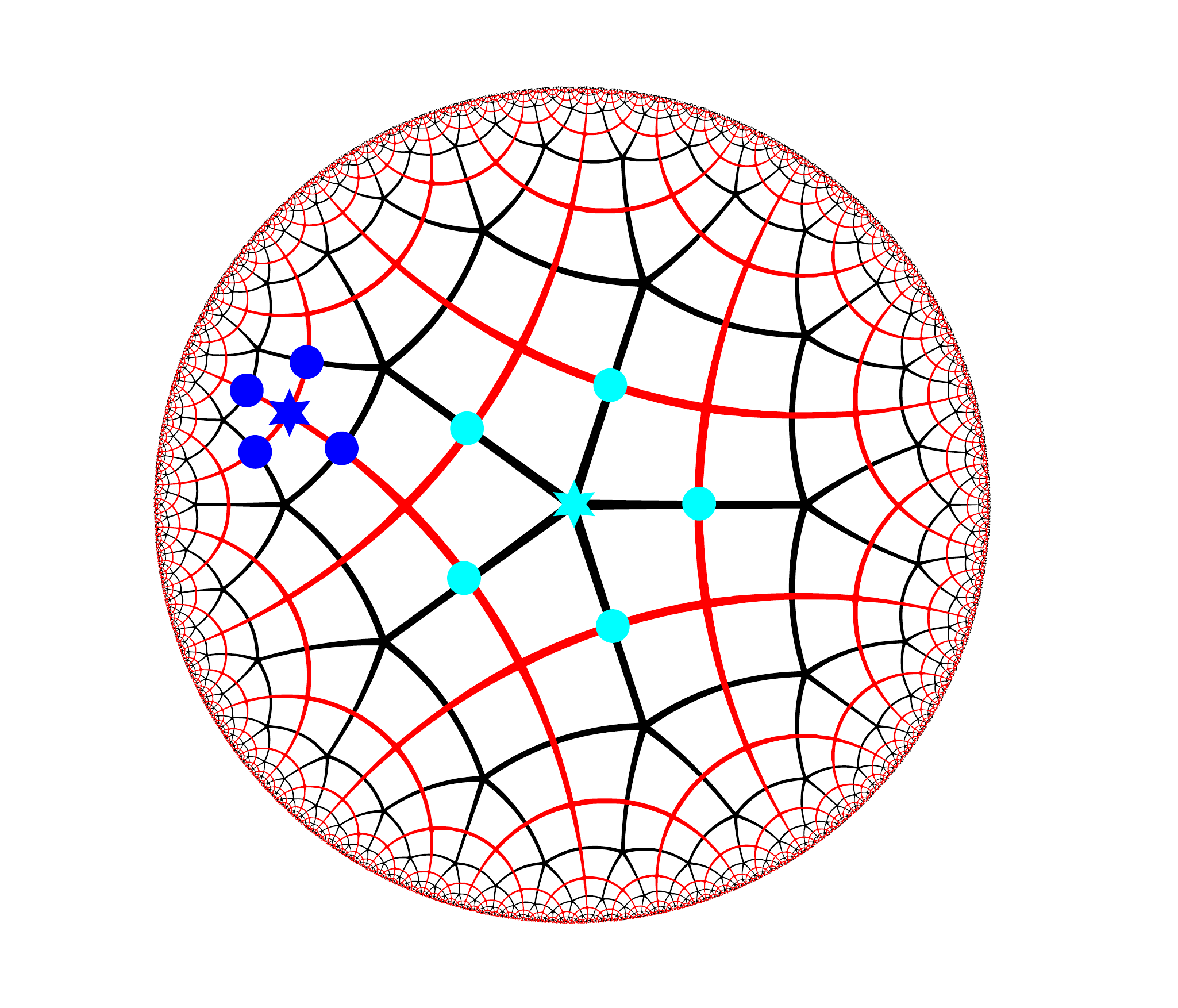}
    \caption{Cluster state connectivity of the foliated hyperbolic surface codes defined on the \{4,5\} tiling, plotted in the Poincaré disk. Black and red lines represent the primal and dual lattices, respectively. Highlighted in dark blue are 4 data qubits (circles) with their corresponding Z--check qubit in the middle (star). Similarly, light blue indicates a primal lattice X--check qubit connected to its 5 neighbouring data qubits. Note that there are connections in the time direction (\emph{i.e.} perpendicular to the page) between primal and dual layers at the data qubit sites.}
    \label{fig:hyperbolic_sc_tiling}
\end{figure}

\subsection{Hyperbolic surface codes}

The high fault-tolerant threshold of the surface code makes it one of the most appealing choices of QEC codes. However, the number of encoded logical qubits per physical qubit (the \emph{encoding rate}) of the surface code approaches zero the distance increases. More precisely, we use the following definition of the encoding rate, $r$:
\begin{equation*}
    r = \frac{k}{n + c},
\end{equation*}
where $k$ is the number of logical qubits, $n$ is the number of data qubits, and $c$ is the number of ancillary check qubits. An asymptotically vanishing encoding rate is common to all QEC codes defined on a 2D tiling of an Euclidean surface~\cite{Breuckmann_2016}. Hyperbolic surface codes are homological stabilizer codes which exploit the properties of hyperbolic geometry to encode more logical qubits than their Euclidean counterparts for a given number of physical qubits~\cite{Breuckmann_2017}. Fig.~\ref{fig:hyperbolic_sc_tiling} shows the Poincaré disk along with the connectivity of the cluster state for a foliated hyperbolic surface code defined on the \{4,5\} tiling. We can see that primal (dual) check nodes have valence of 5 (4). Data qubit nodes have valence 4 in all foliation layers except for the first and last layers, where they have valence of 3.

 The hyperbolic surface codes defined on the \{4,5\} tiling with the lowest number of physical qubits for a given code distance were selected for our simulations. These codes are referred to as \emph{extremal}~\cite{breuckmann2018phdthesishomologicalquantum}, and their parameters were obtained from Table 3.1 in Ref.~\cite{breuckmann2018phdthesishomologicalquantum}. Note that there is an error in the parameters of the distance-12 extremal code there; we use the correct parameters [[4800, 482, 12]]~\cite{Breuckmann_personal}.

\begin{figure}
    \centering
    \includegraphics[width=.45\textwidth]{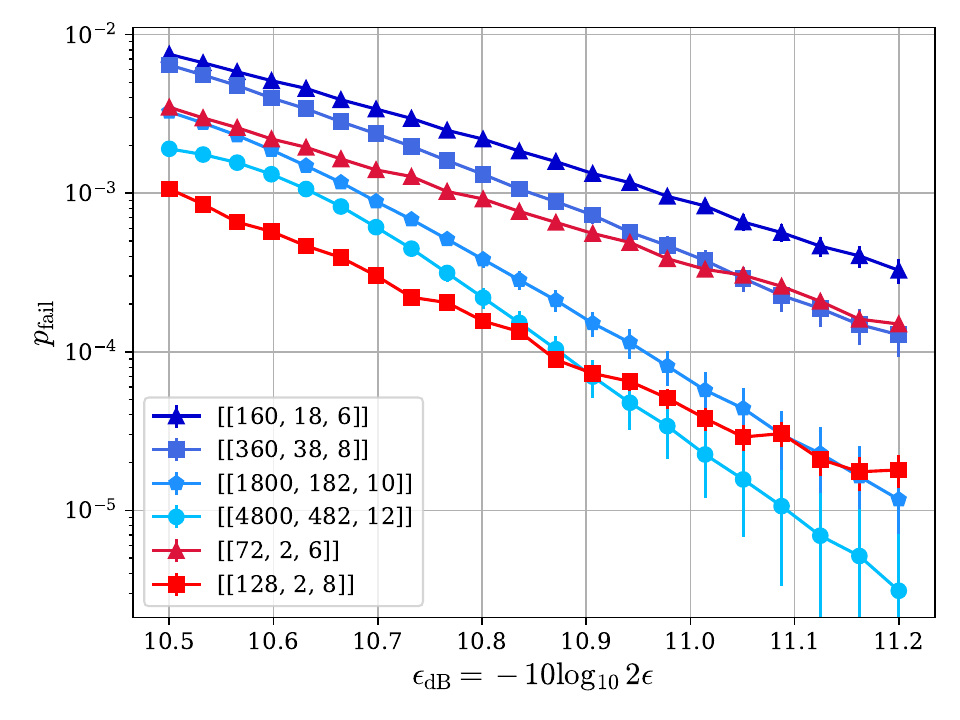}
    \caption{Logical error rates for a fault-tolerant memory under uniform, isotropic GRN with the physical noise parameter $\epsilon$. Blue curves correspond to hyperbolic surface codes defined on the \{4, 5\} tiling and red curves to Euclidean surface codes defined on the \{4, 4\} tiling. The foliated graph state is implemented through the stitching protocol described in the text using a cascade splitter.
    For the hyperbolic codes with $d \in \{6,8\}$ ($d\in\{10,12\})$ each data point represents 100,000 (200,000) Monte Carlo trials.
    For the Euclidean codes, each data point represents $10^6$ trials. The plotted error rates represent the fraction of  trials in which at least one logical error occurred, divided by the number of logical qubits. Correlation-aware inner decoding was used in all simulations to translate homodyne measurements to bit values.}
    \label{fig:hyperbolic_vs_rhg}
\end{figure}

Simulations to obtain Fig.~\ref{subfig:hyperbolic} in the main text proceed closely to those discussed in the Appendix of Ref.~\cite{tzitrin2021fault}, which we now summarize. All simulations were performed using a private version of \textsf{FlamingPy}~\cite{FlamingPy} on the Niagara supercomputer at the SciNet HPC Consortium~\cite{Ponce2019niagara, Loken2010scinet}. Hyperbolic surface codes of various sizes defined on the {4, 5} tiling are instantiated via check matrices and foliated through the procedure described in~\cite{Bolt2016}. The check matrices themselves were obtained from~\cite{CodingTheory}. An updated graph with $M$ modes is generated where the edges of the foliated codes are replaced with dumbbells. Then, an entangling symplectic matrix corresponding to the dumbbells and linear splitters at each macronode is generated. The shift matrix from above is also generated. In each Monte Carlo trial, a vector corresponding to the locations of the Gaussians of two peaks in a GKP $\Ket{\varnothing}$ states is instantiated (\emph{i.e.} a vector of length $2N$ whose entries are randomly chosen between 0 and $\sqrt{2\pi}$). The vector is updated via the symplectic matrices, and combined with the samples from $M$ independent Gaussian distributions to form the raw simulated homodyne measurement outcomes. Those outcomes are processed via the shift matrix and binned via a correlation-aware inner decoder~\cite{Xanadu2024}. Then, the syndrome is computed along with accompanying phase error probabilities and fed into a minimum-weight-perfect matching (MWPM) decoder.

Fig.~\ref{fig:hyperbolic_vs_rhg} shows that extremal hyperbolic surface codes defined on the \{4,5\} tiling (blue curves), which provide substantially more logical qubits than Euclidean surface codes (red curves) for a given code distance, also have a lower error rate in some noise regime which varies for each distance. More specifically, the hyperbolic surface codes in the plot exhibit lower error rates than the simulated Euclidean surface codes at the noise values which lie to the right of their intersection. Error rates in the plot represent the fraction of Monte Carlo trials in which at least one logical error occurred divided by the number of logical qubits of each code. The simulated Euclidean surface codes all had periodic boundaries and were defined on the ``non-rotated'' lattice.
We found that their logical error rates are lower (when comparing fixed distances) than the codes defined on the rotated surface code lattice, which is expected in light of the results in Ref.~\cite{Beverland_2019}.

One can calculate the number of physical qubits required for obtaining the same number of logical qubits and the same error rate as one of the hyperbolic codes with an Euclidean surface code. For instance, the encoding rate of the [[128, 2, 8]] code (Euclidean) is $1/128$. To get the same number of logical qubits as the [[4800, 482, 12]] code (hyperbolic), one would need $482 \times 128 = 61696$ physical qubits, which is an order of magnitude higher than the number of physical qubits of the hyperbolic surface code. These two codes have the same error rates at $\sim10.9$ dB, which implies that, for noise values higher than 10.9 dB, the hyperbolic surface code with parameters [[4800, 482, 12]] has lower error rates than the Euclidean surface code with parameters [[128, 2, 8]] while encoding 241 times more logical qubits.

\vspace*{0.5cm}
\subsection{Bivariate bicycle codes}

Bivariate bicycle codes are QEC codes based on bivariate polynomials, introduced in~\cite{Bravyi2024} as generalizations of bicycle codes~\cite{MacKay2004, Kovalevl2013, Panteleev2021, Panteleev2022bicycle}. They are similar to 2D toric codes, however their checks are not geometrically local and act on six qubits. We consider four of the codes in Ref.~\cite{Bravyi2024}, using parameters [[72, 12, 6]], [[90, 8, 10]], [[144, 12, 12]], and [[288, 12, 18]]. We run the same simulations as we did for hyperbolic surface codes above, but using belief propagation (BP) with ordered statistical decoding (OSD)~\cite{Panteleev2021, Roffe2020} instead of MWPM. The results are encompassed in Fig.~\ref{fig:hyperbolic_sc}b of the main text.

\bibliographystyle{IEEEtran}
%\typeout{}
\bibliography{refs}
%\section*{References}
%\printbibliography[heading=none]

\end{document}